\documentclass[a4paper]{article}
\usepackage{epsfig}
\newcommand{\e}{\mathrm{e}}
\newcommand{\ep}{\epsilon}
\newcommand{\vev}[1]{\left\langle #1 \right\rangle}
\newcommand{\V}{\mathcal{V}}
\newcommand{\ffbox}[1]{\fbox{$\displaystyle #1$}}
\newcommand{\fmslash}[1]{\hbox{$#1$\kern-0.5em\raise0.3ex\hbox{/}}}
\newcommand{\Sp}{\mathrm{Sp}~}
\newcommand{\eq}[1]{$\displaystyle #1$}
\newcommand{\K}[1]{K \left( \frac{#1}{\mu \e^{-t}} \right)}
\newcommand{\D}[1]{\Delta \left( \frac{#1}{\mu \e^{-t}} \right)}
\begin{document}

\title{Wilson's Renormalization Group \\and Its Applications in
  Perturbation Theory\footnote{Lectures given at the APCTP Field
  Theory Winter School, Feb. 2-6, 2006 in Pohang, Korea; preprint
  KOBE-TH-06-02}} \author{H.~Sonoda\footnote{E-mail:
  \texttt{hsonoda@kobe-u.ac.jp}}\\ \textit{Physics Department, Kobe
  University, Kobe 657-8501, Japan}} \date{February 2006}

\maketitle

\begin{abstract}
The general prescription for constructing the continuum limit of a
field theory is explained using Wilson's renormalization group.  We
then formulate the renormalization group in perturbation theory and
apply it to $4$ dimensional $\phi^4$ and QED.
\end{abstract}

\tableofcontents

\newpage
\setcounter{section}{-1}
\section{Introduction}

Quantum Field Theory has a very long history.  It started with the
quantization of the electromagnetic field by Dirac in
1927.\cite{Dirac:1927}  This is almost 80 years ago.  Currently every
student of particle theory (and string theory) takes a course on
quantum field theory which covers such important topics as
perturbative renormalization, gauge theories, spontaneous symmetry
breaking, and the Higgs mechanism.  I assume that you know at least
perturbative renormalization.

The problem with \textbf{perturbative} renormalization theory is the
lack of physical insights.  It consists of procedures for
\textbf{subtracting} UV divergences to get finite results.
Unfortunately, many if not most people still regard renormalization
this way.  It is the purpose of the next four lectures to introduce
the \textbf{physics} of renormalization.

The essence of modern renormalization theory has been known for a long
time.  It was initiated mainly by Ken Wilson in the late 60's.  You
must have heard the expressions such as \textbf{renormalization group
(RG)}, \textbf{RG flows}, \textbf{fixed points}, \textbf{relevant} and
\textbf{irrelevant} parameters.  Unfortunately there is no textbook
introducing these ideas of Wilson's.  After 40 years, the best
reference is still the two lecture notes \cite{Wilson:1973jj,
Wilson:1974mb} by himself.  I believe Wilson's renormalization theory
is best studied in a second course on field theory, and my four
lectures will give merely an outline.

The four lectures are organized as follows.  In lecture 1, we start
with concrete examples of renormalization to introduce the relation
between criticality and renormalizability.  We will give hardly any
derivation, but the examples will illuminate the meaning of
renormalization.  In lecture 2, we introduce the \textbf{exact
renormalization group (ERG)} as a tool to understand the nature of
renormalization.  The emphasis is on the ideas of fixed points,
relevance/irrelevance of parameters, and universality.  In lectures 3
\& 4, we apply Wilson's exact renormalization group to perturbation
theory.  This was initiated by J.~Polchinski in
1984.\cite{Polchinski:1983gv} We mainly discuss the $\phi^4$ theory in
$4$ dimensions in lecture 3, and QED in lecture 4.  I think that it is
a reflection of the depth of the renormalization group that after
almost 40 years of conception it is still under active
research.\cite{RG2005}

\section{Lecture 1 -- Continuum Limits}

The purpose of the first lecture is to familiarize ourselves with the
concept of renormalization through concrete examples.  Before we
start, we should agree on the use of the euclidean metric as opposed
to the Minkowski metric.  
\[
\eta_{\mu\nu} = \mathrm{diag} (1,-1,-1,-1) \longrightarrow
\delta_{\mu\nu} = \mathrm{diag} (1,1,1,1)
\]
Given an $n$-point Green function of a scalar field $\phi$
\[
\vev{\phi (\vec{x}_1, x_1^0) \cdots \phi (\vec{x}_n, x_n^0)}
\]
we obtain an $n$-point correlation function
\[
\vev{\phi (x_1) \cdots \phi (x_n)}
\]
by the analytic continuation
\[
x_i^0 \longrightarrow - i x_i^4\qquad (i=1,\cdots,n)
\]
For example, the free scalar propagator
\[
\int \frac{d^D p}{(2 \pi)^D} \frac{i ~\e^{- i p x}}{p^2 - m^2 + i \ep}
\qquad (p x \equiv p^0 x^0 - \vec{p} \cdot \vec{x})
\]
becomes
\[
\int \frac{d^D p}{(2 \pi)^D} \frac{\e^{i p x}}{p^2 + m^2}\qquad
(p x \equiv p_4 x_4 + \vec{p} \cdot \vec{x})
\]
\textbf{Problem 1-1}: Derive this.\\
Hence, the free propagator in Minkowski space
\[
\frac{i}{p^2 - m^2 + i \ep}
\]
becomes
\[
\frac{1}{p^2 + m^2}
\]
in euclidean space.

\subsection{The idea of a continuum limit}

The idea of a continuum limit is very simple and must be already
familiar to you.  We consider a theory with a momentum cutoff
$\Lambda$, meaning that the theory is defined only up to the scale
$\Lambda$.  For example, a lattice theory defined on a cubic lattice
of a lattice unit
\[
a = \frac{1}{\Lambda}
\]
has the momentum cutoff $\Lambda$.  (Figure 1)
\begin{figure}
\begin{center}
\epsfig{file=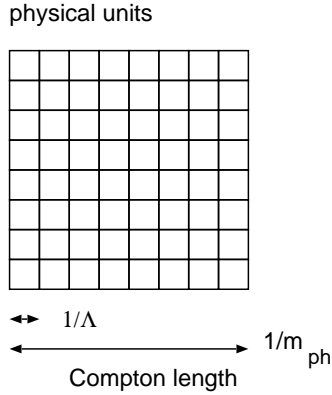}
\caption{A cubic lattice in physical units}
\end{center}
\end{figure}

\textbf{The continuum limit} is the limit
\[
\Lambda \to \infty
\]
and \textbf{renormalization} is the specific way of taking the continuum limit
so that the physical mass scale $m_{\mathrm{ph}}$, say the mass of an
elementary particle, remains finite.

From the viewpoint of a lattice theory, it is more natural to measure
distances in lattice units.  Hence, the lattice unit becomes simply
$1$.
\begin{figure}
\begin{center}
\epsfig{file=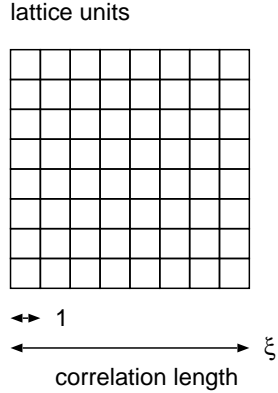}
\caption{A cubic lattice in lattice units}
\end{center}
\end{figure}
In this convention, the Compton length or equivalently the inverse of
the physical mass is a dimensionless number $\xi$ called the
correlation length.  Therefore, we obtain
\[
\xi \, a = \frac{1}{m_{\mathrm{ph}}} \Longrightarrow
\ffbox{\xi = \frac{\Lambda}{m_{\mathrm{ph}}}}
\]
Clearly, $\xi \to \infty$ as we take $\Lambda \to \infty$ while
keeping $m_{\mathrm{ph}}$ finite.  Thus, as we take the continuum
limit, the lattice theory must obtain an infinite correlation length.

A lattice theory with an infinite correlation length is called a
\textbf{critical} theory.  Therefore, \textbf{to obtain a continuum
limit the corresponding lattice theory must be critical}.  Let us look
at examples.\footnote{ See Appendix A for an example of asymptotic
free theories.  In lecture 2 we simply quote the results without any
derivation. See Appendix B for explicit calculations for $\phi^4$ in
$3$ \& $4$ dimensions.}

\subsection{Ising model in $2$ dimensions}

The Ising model on a square lattice is defined by the action
\[
S = - K \sum_{\vec{n} = (n_1,n_2)} \sum_{i=1}^2 \sigma_{\vec{n}}
  \sigma_{\vec{n} + \hat{i}}
\]
At each site $\vec{n}$ of the lattice, we introduce a classical spin
variable $\sigma_{\vec{n}} = \pm 1$.  The parameter $K$ is a
dimensionless positive constant, which we can regard as the inverse of
a reduced (i.e., dimensionless) temperature.
\[
K \sim \frac{1}{T}
\]
The partition function is defined by
\[
Z (K) = \sum_{\sigma_{\vec{n}} = \pm 1} \e^{- S}
\]
and the correlation functions are defined by
\[
\vev{\sigma_{\vec{n}_1} \cdots \sigma_{\vec{n}_N}}_K =
\frac{\sum_{\sigma = \pm 1} \sigma_{\vec{n}_1} \cdots
\sigma_{\vec{n}_N} \e^{- S}}{Z (K)}
\]

The action is invariant under the global $\mathbf{Z_2}$ transformation
\[
(\forall \vec{n})\quad \sigma_{\vec{n}} \longrightarrow -
\sigma_{\vec{n}}
\]  With respect to this symmetry, the model has two phases:
\begin{itemize}
\item High temperature phase $K < K_c$: the $\mathbf{Z}_2$ symmetry is
exact, and
\[
\vev{\sigma_{\vec{n}}} = 0
\]
\item Low temperature phase $K > K_c$: the $\mathbf{Z}_2$ symmetry is
  spontaneously broken, and
\[
\vev{\sigma_{\vec{n}}} = s (K) \ne 0
\]
\end{itemize}
For large $|\vec{n}|$, the two-point function behaves exponentially as
\[
\vev{\sigma_{\vec{n}} \sigma_{\vec{0}}}_K \sim \e^{-
  \frac{|\vec{n}|}{\xi}}
\]
This defines the correlation length $\xi (K)$.  At $K = K_c$ the
theory is critical with $\xi = \infty$.  Two \textbf{critical
exponents}
\[
\ffbox{y_E = 1} \quad \textrm{and} \quad
\ffbox{x_h = \frac{1}{8}}
\]
characterize the theory near criticality as follows:
\begin{itemize}
\item As $K \to K_c$, the correlation length behaves as
\[
\xi \sim |K - K_c|^{- \frac{1}{y_E}} = \frac{1}{|K - K_c|}
\]
\item As $K \to K_c + 0$, the VEV behaves as
\[
s \sim |K-K_c|^{\frac{h}{y_E}} = |K-K_c|^{\frac{1}{8}}
\]
\end{itemize}
\begin{figure}
\begin{center}
\epsfig{file=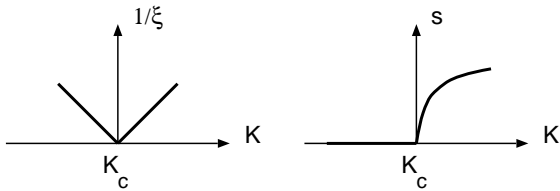}
\caption{The critical exponents $y_E, x_h$ characterize the
  correlation length $\xi$ and VEV $s$ near the critical point $K =
  K_c$.}
\end{center}
\end{figure}
The correlation functions near the critical point $K \simeq K_c$ obeys
\textbf{the scaling law}:
\[
\ffbox{ \vev{\sigma_{\vec{n}_1} \cdots \sigma_{\vec{n}_N}}_K \simeq |K
  - K_c|^{N \frac{x_h}{y_E}} F_N^{\pm} \left( \frac{\vec{n}_1 -
    \vec{n}_N}{\xi}, \cdots, \frac{\vec{n}_{N-1} - \vec{n}_N}{\xi}
  \right)}
\]
where $\pm$ for $K > (<) K_c$.  For $N > 1$, the scaling law is valid
  only for large separation of lattice sites:
\[
|\vec{n}_i - \vec{n}_j| \gg 1 \quad (i \ne j)
\]

For $N=1$, the scaling law simply boils down to the exponential
behavior of $s(K)$ near criticality.  For $N=2$, the scaling law gives
\[
\vev{\sigma_{\vec{n}} \sigma_{\vec{0}}}_K
 \simeq |K-K_c|^{\frac{1}{4}} F_2^\pm \left(\frac{\vec{n}}{\xi} \right)
\]
where $\xi \sim \frac{1}{|K-K_c|}$.  For the limit to exist as $K \to
K_c$, the function $F_2^\pm$ must behave like
\[
F_2^\pm (x) \sim x^{- 1/4}
\]
for $x \ll 1$.  Hence, at the critical point $K=K_c$, the two-point
function is given by the exponential
\[
\vev{\sigma_{\vec{n}} \sigma_{\vec{0}}}_{K_c} \sim
 \frac{1}{|\vec{n}|^{2 x_h}} = \frac{1}{|\vec{n}|^\frac{1}{4}}\quad
 (|\vec{n}| \gg 1)
\]
In fact this is another way of introducing the critical exponent
$x_h$.

The scaling law introduced above implies that we can renormalize the
Ising model to construct a scalar field theory as follows:
\[
\ffbox{
\vev{\phi (\vec{r}_1) \cdots \phi (\vec{r}_{N})}_{m; \mu}
 \equiv \lim_{t \to \infty} \e^{\frac{N t}{8}} \vev{\sigma_{\vec{n}_1
 = \mu \vec{r}_1 \e^t} \cdots \sigma_{\vec{n}_N = \mu \vec{r}_N
 \e^t}}_{K = K_c - \frac{m}{\mu} \e^{-t}}
}
\]
where both $m$ and $\mu$ have mass dimension $1$.\footnote{We chose a
  sign convention so that the $\mathbf{Z}_2$ is spontaneously broken
  for $m < 0$.}

Let us explain this formula in several steps:
\begin{enumerate}
\item $K = K_c - \frac{m}{\mu} \e^{-t}$ --- Hence, as $t \to \infty$,
  the theory approaches criticality.  The particular $t$ dependence
  was chosen so that $\xi \propto \e^t$.
\item Necessity of $\mu$ --- $\mu$ was introduced so that the physical
  length of a lattice unit is $\frac{1}{\mu} \e^{-t}$.  Hence,
  $\vec{r} = \vec{n} \frac{1}{\mu} \e^{-t}$ has mass dimension $-1$,
  and $m$ has mass dimension $1$.
\item Given an arbitrary coordinate $\vec{r}$, $\vec{n} = \mu \vec{r}
  \e^t$ is not necessarily a vector with integer components.  Since
  $\e^t \gg 1$, however, we can always find an integral vector
  $\vec{n}$ which approximates $\mu \vec{r} \e^t$ to the accuracy
  $\e^{-t} \ll 1$.
\item Applying the scaling law, we can compute the limit as
\begin{eqnarray*}
&&\vev{\phi (\vec{r}_1) \cdots \phi (\vec{r}_{N})}_{m; \mu} \\
&=&
\lim_{t \to \infty} \e^{\frac{N t}{8}} |K - K_c|^{\frac{N}{8}}
F_N^\pm \left( \frac{\mu \e^t |K-K_c|}{c_\pm} (\vec{r}_1 -
\vec{r}_N), \cdots \right)\\
&=& \left(\frac{m}{\mu}\right)^{\frac{N}{8}} F_N^\pm \left(
\frac{m}{c_\pm} (\vec{r}_1 - \vec{r}_N), \cdots \right)
\end{eqnarray*}
where we used
\[
\xi \simeq \frac{c_\pm}{|K-K_c|}
\]
Thus, the limit exists.  The limit depends not only on the mass
parameter $m$ but also on the arbitrary mass scale $\mu$.
\item \textbf{RG equation} --- The correlation function satisfies
\[
\vev{\phi (\e^{- \Delta t} \vec{r}_1) \cdots \phi (\e^{- \Delta t}
\vec{r}_{N})}_{m \e^t; \mu} = \e^{N \frac{\Delta t}{8}} \vev{\phi
(\vec{r}_1) \cdots \phi (\vec{r}_{N})}_{m; \mu}
\]
This implies that 
 the scale change of the coordinates
\[
(\forall i)\quad \vec{r}_i \longrightarrow \vec{r}_i \e^{- \Delta t}
\]
can be compensated by the change of the mass parameter $m$:
\[
m \longrightarrow m \e^{\Delta t}
\]
and renormalization of the field:
\[
\phi \longrightarrow \e^{\frac{\Delta t}{8}} \phi
\]
Hence, $y_E = 1$ is the \textbf{scale dimension} of $m$, and $x_h =
\frac{1}{8}$ is that of $\phi$. The general solution of the RG
equation is given by the scaling formula with $F_N^\pm$ as arbitrary
functions.
\item $\mu$ dependence (alternative RG equation) 
\[
\vev{\phi (\vec{r}_1) \cdots \phi (\vec{r}_{N})}_{m; \mu \e^{- \Delta t}}
= \e^{\frac{\Delta t}{8}} \vev{\phi
(\vec{r}_1) \cdots \phi (\vec{r}_{N})}_{m; \mu}
\]
This is obtained from the previous RG equation by dimensional
analysis.  The change of $\mu$ is compensated by renormalization of
$\phi$.  
\item Elimination of $\mu$ --- If we want, we can eliminate the
  arbitrary scale $\mu$ from the continuum limit by giving the mass
  dimension $\frac{1}{8}$ to the scalar field.  By writing
  $\mu^{\frac{1}{8}} \phi$ as the new scalar field $\phi$, we obtain
\[
\vev{\phi (\vec{r}_1) \cdots \phi (\vec{r}_{N})}_m
 = m^{\frac{N}{8}} F_N^\pm \left(
\frac{m}{c_\pm} (\vec{r}_1 - \vec{r}_N), \cdots \right)
\]
\end{enumerate}

Before ending, let us examine the short distance behavior using RG.
The RG equation can be rewritten as
\[
\vev{\phi (\e^{- t} \vec{r}_1) \cdots \phi (\e^{- t} \vec{r}_N)}_{m;
  \mu} = \e^{N \frac{t}{8}} \vev{ \phi (\vec{r}_1) \cdots \phi
  (\vec{r}_N)}_{m \e^{- t}; \mu}
\]
Hence, in the short distance limit, the correlation functions are
given by those at the critical point $m=0$:
\[
\vev{\phi (\e^{- t} \vec{r}_1) \cdots \phi (\e^{- t} \vec{r}_N)}_{m;
  \mu} \stackrel{t \gg 1}{\simeq} \e^{N \frac{t}{8}} \vev{\phi
  (\vec{r}_1) \cdots \phi (\vec{r}_N)}_{0; \mu}
\]
Especially for the two-point function, we obtain
\[
\vev{\phi (\e^{-t} \vec{r}) \phi (\vec{0})}_{m; \mu} \stackrel{t \gg
  1}{\simeq} 
\frac{\mathrm{const}}{(\mu r \e^{- t})^{\frac{1}{4}}}
\]

\subsection{Ising model in $3$ dimensions}

We can construct the continuum limit of the $3$ dimensional Ising
model the same way as for the $2$ dimensional one.  The only
difference is the value of the critical exponents $y_E$ and $x_h$.
\[
y_E \simeq 1.6,\quad \eta \equiv 2 x_h - 1 \simeq 0.04
\]
These are known only approximately.  $\eta$ gives the difference of
  the scale dimension of the scalar field from the free field value,
  and is called \textbf{the anomalous dimension}.

In defining the continuum limit, we can give any engineering dimension
to the scalar field.  Here, let us pick $\frac{1}{2}$, the same as the
free scalar field.  The two-point function can be defined as
\[
\vev{\phi  (\vec{r}) \phi (\vec{0})}_{g; \mu}
\equiv \mu \lim_{t \to \infty} \e^{(1 + \eta) t} \vev{\sigma_{\vec{n}
    = \mu \vec{r} \e^t} \sigma_{\vec{0}}}_{K = K_c -
  \frac{g}{\mu^2} \e^{- y_E t}}
\]
\textbf{Problem 1-2}: Define the $n$-point function.\\
We have arbitrarily given the engineering $2$ to the parameter $g$.

The two-point function obeys the following RG equation:
\[
\vev{\phi (\vec{r} \e^{- \Delta t}) \phi (\vec{0})}_{g \e^{y_E \Delta
 t}; \mu} = \e^{(1+\eta) \Delta t} \vev{\phi (\vec{r}) \phi
 (\vec{0})}_{g; \mu}
\]
This implies the short-distance behavior
\[
\vev{\phi (\vec{r} \e^{-t}) \phi (\vec{0})}_{g; \mu}
\stackrel{t \gg 1}{\simeq} \mathrm{const} \frac{\mu}{(\mu r
  \e^{-t})^{1 + \eta}}
\]

\subsection{Alternative: $\phi^4$ on a cubic lattice}

The same continuum limit can be obtained from a different model.  Let
$\phi_{\vec{n}}$ be a real variable taking a value from $- \infty$ to
$\infty$.  We consider the $\phi^4$ theory on a cubic lattice:
\[
S = \sum_{\vec{n}} \left[
\frac{1}{2} \sum_{i=1}^3 \left( \phi_{\vec{n}+\hat{i}} -
\phi_{\vec{n}} \right)^2 + \frac{m_0^2}{2} \phi_{\vec{n}}^2 +
\frac{\lambda_0}{4!} \phi_{\vec{n}}^4 \right]
\]
where $\lambda_0 > 0$, but $m_0^2$ can be negative.\footnote{See
Appendix B for explicit calculations.}

For $\lambda_0$ fixed, the theory has two phases depending on the
value of $m_0^2$:
\begin{itemize}
\item \textbf{symmetric phase} $m_0^2 > m_{0, cr}^2 (\lambda_0)$:
  $\mathbf{Z}_2$ is intact.
\item \textbf{broken phase} $m_0^2 < m_{0, cr}^2 (\lambda_0)$:
  $\mathbf{Z}_2$ is spontaneously broken, and $\vev{\phi_{\vec{n}}}
  \ne 0$.
\end{itemize}
Note that the critical value $m_{0, cr}^2 (\lambda_0)$ depends on
$\lambda_0$.

The continuum limit is obtained as
\[
\vev{\phi (\vec{r}) \phi (\vec{0})}_{g; \mu} \equiv \mu \lim_{t \to
\infty} \e^{ (1+\eta) t} \vev{\phi_{\vec{n} = \mu \vec{r} \e^t}
\phi_{\vec{0}}}_{m_0^2 = m_{0, cr}^2 (\lambda_0) + \e^{- y_E t}
\frac{g}{\mu^2}}
\]
where $y_E, \eta$ are the same critical exponents as in the Ising
model.  This limit is not necessarily independent of $\lambda_0$.  For
independence, we need to do rescale both $\phi$ and $g$:
\[
\left\lbrace\begin{array}{c@{~\longrightarrow~}c}
 \phi & \sqrt{z (\lambda_0)} ~\phi\\
 g & z_m (\lambda_0)~g
\end{array}
\right.
\]
and define
\[
\vev{\phi (\vec{r}) \phi (\vec{0})}_{g; \mu} \equiv z (\lambda_0) \mu
\lim_{t \to \infty} \e^{ (1+\eta) t} \vev{\phi_{\vec{n} = \mu \vec{r}
\e^t} \phi_{\vec{0}}}_{m_0^2 = m_{0, cr}^2 (\lambda_0) + \e^{- y_E t}
z_m (\lambda_0) \frac{g}{\mu^2}}
\]
\textbf{Universality} consists of two statements:
\begin{enumerate}
\item $y_E, \eta$ are the same as in the Ising model.
\item The continuum limit is the same as in the Ising model.  (We only
  have to choose $z(\lambda_0)$ and $z_m (\lambda_0)$ appropriately.)
\end{enumerate}

We have defined the continuum limit using the lattice units for the
lattice theory.  How do we take the continuum limit if we use the
physical units instead?  To go to the physical units, we assign the length
\[
a = \frac{1}{\Lambda} = \frac{1}{\mu e^t}
\]
to the lattice unit.  The action is now given by
\[
S = \underbrace{a^3 \sum_{\vec{n}}}_{= \int d^3 r} \Bigg[ \frac{1}{2}
  \sum_{i=1}^3 \underbrace{\frac{1}{a^2} \left( \varphi_{\vec{r} + a
  \hat{i}} - \varphi_{\vec{r}} \right)^2}_{= (\partial_i \varphi)^2} +
  \frac{m_{bare}^2}{2} \varphi_{\vec{r}}^2 + \frac{\lambda_{bare}}{4!}
  \varphi_{\vec{r}}^4 \Bigg]
\]
where
\[
\left\lbrace\begin{array}{c@{~\equiv~}l}
 \vec{r} & \vec{n} a = \frac{\vec{n}}{\Lambda}\\
 \varphi_{\vec{r}} & \frac{1}{\sqrt{a}} \phi_{\vec{n}} =
 \sqrt{\Lambda} \phi_{\vec{n}}\\
 m_{bare}^2 & \frac{m_0^2}{a^2} = m_0^2 \Lambda^2\\
 \lambda_{bare} & \frac{\lambda_0}{a} = \lambda_0 \Lambda
\end{array}\right.
\]
Then, to obtain the continuum limit we must choose
\[
\left\lbrace\begin{array}{c@{~=~}l} m_{bare}^2 & \Lambda^2 m_{0,cr}^2
 (\lambda_0) + z_m (\lambda_0) g \left( \frac{\Lambda}{\mu} \right)^{2
 - y_E}\\ \lambda_{bare} & \Lambda \lambda_0
\end{array}\right.
\]
and we obtain
\[
\vev{\phi (\vec{r}) \phi (\vec{0})}_{g; \mu}
 = z (\lambda_0) \lim_{\Lambda \to \infty}
 \left( \frac{\Lambda}{\mu} \right)^\eta \vev{ \varphi_{\vec{r}}
 \varphi_{\vec{0}} }_{m_{bare}^2, \lambda_{bare}}
\]
Note that $\lambda_0 > 0$ is a finite arbitrary constant.  The bare
squared mass has not only a quadratic divergence but also a divergence
of power $2 - y_E \simeq 0.4$.  The bare coupling is linearly
divergent.  We see clearly that \textbf{the UV divergences of
parameters are due to the use of physical units.}  If we use the
lattice units, there is no divergence.\footnote{Except for the
divergences in the lattice distance $\vec{n}$ and the normalization
constant $\e^{(1+\eta) t}$.}

\subsection{Ising model in $4$ dimensions}

The $4$ dimensional case is very different from the lower dimensional
cases in that we cannot take the continuum limit $\Lambda \to \infty$
without obtaining a free theory.  This is called
\textbf{triviality}.  For example, we obtain
\begin{eqnarray*}
\vev{\phi  (\vec{r}) \phi (\vec{0})}_{m^2; \mu}
&\equiv& z \mu^2 \lim_{t \to \infty} \e^{2 t} \vev{\sigma_{\vec{n}
    = \mu \vec{r} \e^t} \sigma_{\vec{0}}}_{K = K_c -
  z_m \frac{m^2}{\mu^2} \e^{- 2 t} t^{- \beta_m}}\\
&=& \int_p \frac{\e^{i p r}}{p^2 + m^2}
\end{eqnarray*}
where
\[
\beta_m = - \frac{1}{3}
\]
and $z, z_m$ are appropriate constants.\footnote{See Appendix B for
explicit calculations using the $4$ dimensional $\phi^4$ theory.}

To keep the theory interacting, we must choose
\[
\Lambda = \mu \e^t
\]
large but finite so that we can define a coupling constant
$\lambda$ by\footnote{This $\lambda$ is what we usually call $\frac{3
    \lambda}{(4 \pi)^2}$} 
\[
\e^t = \e^{\frac{1}{\lambda}} \left(\frac{\lambda}{1 - c
  \lambda}\right)^{- c}
\]
where\footnote{To follow the rest, we don't lose much by assuming
$c=0$.}
\[
c = \frac{17}{27}
\]
$\lambda$ is a function of $t = \ln \frac{\Lambda}{\mu}$ satisfying
the differential equation
\[
\frac{d}{dt} \lambda = - \lambda^2 + c \lambda^3
\]
\textbf{Problem 1-3:} Derive this.\\
Note that for $t \gg 1$, we find
\[
\lambda = \frac{1}{t - c \ln t + \mathrm{O} (1/t)} \ll 1
\]
Hence, \textbf{a large cutoff $\Lambda \gg \mu$ implies a small
coupling}.

We define the $\phi^4$ theory by
\begin{eqnarray*}
\vev{\phi (\vec{r}) \phi (\vec{0})}_{m^2, \lambda; \mu} &\equiv& z
\mu^2 \e^{2 t} \vev{\sigma_{\vec{n} = \mu \vec{r} \e^t}
\sigma_{\vec{0}}}_{K = K_c - z_m \frac{m^2}{\mu^2} \e^{- 2 t}
\left(\frac{\lambda}{1 - c \lambda}\right)^{\beta_m}}\\ &=& z \Lambda^2
\vev{\sigma_{\vec{n} = \Lambda \vec{r}} \sigma_{\vec{0}}}_{K = K_c -
z_m \frac{m^2}{\Lambda^2} \left(\frac{\lambda}{1 - c
\lambda}\right)^{\beta_m}}
\end{eqnarray*}
To derive the RG equation, we scale $\vec{r}$ infinitesimally to
  $\vec{r} \e^{- \Delta t}$.  To keep $\vec{n}$ invariant, we must
  change $t$ to $t + \Delta t$.  This changes $\lambda$ by
\[
\Delta \lambda = \Delta t \left( - \lambda^2 + c \lambda^3 \right)
\]
To keep $K$ invariant, we must change $m^2$ by 
\[
\Delta m^2 = \Delta t \cdot m^2 (2 + \beta_m \lambda)
\]
Hence, we obtain\footnote{There is no anomalous dimension for the
  field $\phi$.  In general it is of order $\lambda^2$, and it can be
  eliminated by field redefinition.}
\[
\vev{\phi (\vec{r} \e^{- \Delta t}) \phi (\vec{0})}_{m^2 + \Delta m^2,
  \lambda + \Delta \lambda; \mu} = \e^{2 \Delta t} \vev{\phi (\vec{r})
  \phi (\vec{0})}_{m^2, \lambda; \mu}
\]
In perturbation theory we compute the correlation functions in powers
of $\lambda$.  Since the cutoff is given by
\[
\Lambda = \mu \e^t = \mu \e^{\frac{1}{\lambda}} \left(
\frac{\lambda}{1 - c \lambda} \right)^{-c} \stackrel{\lambda \to
  0}{\longrightarrow} + \infty
\]
it is infinite in perturbation theory.  Hence, \textbf{as long as we
use perturbation theory, we can take the continuum limit of the
$\phi^4$ theory}.  

\section{Lecture 2 -- Wilson's RG}

There are three important issues with renormalization:
\begin{enumerate}
\item \textbf{how to renormalize a theory} --- in the previous lecture
  we have seen how to take the continuum limit.  We would like to
  understand why the limit exists.
\item \textbf{universality} --- the continuum limit does not depend on
  the specific model we use.  For example, the three dimensional
  $\phi^4$ theory can be defined using either the Ising model or the
  $\phi^4$ theory on a cubic lattice.
\item \textbf{finite number of parameters} --- the continuum limit
  depends only on a finite number of parameters such as $g$ for the
  three dimensional $\phi^4$ and $m^2, \lambda$ for the four
  dimensional $\phi^4$.  We would like to understand why.
\end{enumerate}

Ken Wilson clarified all these by introducing his
\textbf{RG}.\cite{Wilson:1973jj, Wilson:1974mb} We may add the
adjective \textbf{exact} or Wilson's or Wilsonian to distinguish it
from the RG acting on a finite number of parameters.  Exact RG can be
abbreviated as \textbf{ERG}.

\subsection{Definition of ERG} 

It is difficult to formulate ERG precisely.  (We will omit E from ERG
from now on.) In the next lecture we
will see how to formulate it within perturbation theory.  Here, we
simply assume the existence of a non-perturbative formulation, and use
it to illustrate the expected properties of RG.\footnote{For a more
rigorous approach than presented here, please consult with Wegner's
article \cite{Wegner:74}.}  It is important to note that we don't need
to use RG to define continuum limits as we saw in the previous
lecture.  The most important role of RG is to give us an insight into
the actual procedure of renormalization.  Without RG it is hard to
understand the meaning of renormalization.

To simplify our discussion, we introduce only one \textbf{real} scalar
field.  Given a momentum cutoff $\Lambda$, the scalar field can be
expanded as 
\[
\varphi (\vec{r}) = \int_{p < \Lambda} \e^{i p r} \tilde{\varphi}
(\vec{p})
\]
where $\int_p \equiv \int \frac{d^D p}{(2 \pi)^D}$ for $D$-dimensional
space.  We would like to define RG so that the cutoff does not change
under renormalization.  This calls for the use of ``lattice units'':
we make everything dimensionless by multiplying appropriate powers of
$\Lambda$.  Hence, the scalar field can be expanded as
\[
\phi (\vec{r}) = \int_{p < 1} \e^{i p r} \tilde{\phi} (\vec{p})
\]
where $\vec{r}$ is a dimensionless spatial vector in units of
$\frac{1}{\Lambda}$, and $\vec{p}$ is a dimensionless momentum in
units of $\Lambda$.

A theory is defined by the action $S [\phi]$ which is a functional of
$\phi$.\footnote{Precisely speaking this is not quite correct.  Two
different actions related to each other by a change of variables $\phi
\to \Phi (\phi)$ correspond to the same theory.  We will ignore this
subtlety in the rest.  Again consult with \cite{Wegner:74} for more
serious discussions.}  Besides the usual invariance under translation
and rotation, there are two constraints on the choice of $S$:
\begin{enumerate}
\item \textbf{positivity} --- $S$ must be bounded from below so that
  the functional integral
\[
Z = \int [d \phi] \e^{ - S[\phi]}
\]
 is well defined.\footnote{Of course, what is well defined is $- \ln Z$
 by unit volume (free energy/volume).}
\item \textbf{locality} --- If we expand $S$ in powers of $\phi$, the
  interaction kernel
\[
 \frac{\delta^n S}{\delta \phi (\vec{r}_1) \cdots \delta \phi
 (\vec{r}_n)} \Bigg|_{\phi = 0}
\]
 is non-vanishing only within a separation of order $1$.  This
 condition is hard to incorporate, though, unless we express $S$ as a
 power series of $\phi$.
\end{enumerate}
These two properties are very important physically, but they do not
play any major roles in the following discussion of RG.  Hence, you
may happily ignore them ;-)

RG is introduced as a transformation from one $S$ to another.  It
consists of three steps:
\begin{enumerate}
\item \textbf{integration over high momentum modes} --- we integrate
  over $\tilde{\phi} (\vec{p})$ only for
\[
\e^{- \Delta t} < p < 1
\]
 where $\Delta t > 0$ is infinitesimal.  We obtain
\[
\e^{- S' [\phi]} \equiv \int \left[ d \tilde{\phi} (\vec{p})
  \right]_{\e^{- \Delta t} < p < 1} \e^{ - S [\phi]}
\]
 where $S' [\phi]$ depends only on $\tilde{\phi} (\vec{p})$ with
\[
p < \e^{- \Delta t}
\]
\item \textbf{rescaling of space or equivalently momentum} --- we are
  left with the scalar field
\[
\phi (\vec{r}) = \int_{p < \e^{- \Delta t}} \e^{i p r} \tilde{\phi}
(\vec{p})
\]
  We now define
\[
\phi' (\vec{r}) \equiv \e^{\frac{D-2}{2} \Delta t} \phi (\vec{r}
  \e^{\Delta t}) = \e^{- \frac{D+2}{2} \Delta t} \int_{p < 1} \e^{ipr}
  \tilde{\phi} (\vec{p} \e^{- \Delta t})
\]
 so that its Fourier mode is given by
\[
\tilde{\phi}' (\vec{p}) = \e^{- \frac{D+2}{2} \Delta t} \tilde{\phi}
  (\vec{p} \e^{- \Delta t})
\]
 and that the momentum $\vec{p}$ of $\phi'$ ranges over the entire
domain
\[
p < 1
\]
\item \textbf{renormalization of $\phi$} --- we change the
  normalization of the field
\[
\phi'' (\vec{r}) \equiv \left(1 + \Delta t \cdot \left(\gamma -
\frac{D-2}{2}\right)\right) \phi' (\vec{r}) \stackrel{r \gg 1}{\simeq}
\e^{\gamma \Delta t} \phi (\vec{r} \e^{\Delta t})
\]
and denote the action $S' [\phi]$ by $(S + \Delta S) [\phi'']$.  Like
the original $S$, the transformed action $S + \Delta S$ is defined for
the field with momentum $p < 1$.  We determine the renormalization
constant $\gamma$, for example, by the condition that the kinetic
term\footnote{The coefficient of the quadratic term can be expanded in
powers of $p^2$.  The kinetic term is the linear term in this
expansion.}  of $(S+\Delta S) [\phi]$ is normalized as
\[
\frac{1}{2} \int_{p < 1} p^2 \tilde{\phi} (\vec{p}) \tilde{\phi} (-
\vec{p})
\]
The value of $\gamma$ depends on $S$, and we may write $\gamma (S)$
for clarity.  This last step is necessary for the RG transformation to
obtain fixed points.
\end{enumerate}

The transformation from $S$ to $S+\Delta S$ is the infinitesimal RG
transformation $R_{\Delta t}$ so that we write $S + \Delta S =
R_{\Delta t} S$.  Then, under the $R_{\Delta t}$, we obtain
\[
\ffbox{
\vev{\phi (\vec{r}_1) \cdots \phi (\vec{r}_n)}_S
 = (1 - n \gamma \Delta t) \vev{\phi (\vec{r}_1 \e^{- \Delta t}) \cdots
 \phi (\vec{r}_n \e^{- \Delta t})}_{R_{\Delta t} S}
}
\]
This is valid for large separations
\[
|\vec{r}_i - \vec{r}_j| \gg 1
\]
for which the modes $\e^{- \Delta t} < p < 1$ do not contribute.  By
integrating over $R_{\Delta t}$, we obtain a finite RG transformation
$R_t$, under which
\[
\vev{\phi (\vec{r}_1) \cdots \phi (\vec{r}_n)}_S = \e^{ - n \int_0^t
    dt'\, \gamma (R_{t'} S) } \vev{\phi (\e^{-t} \vec{r}_1) \cdots \phi
    (\e^{-t} \vec{r}_n)}_{R_t S}
\]
The RG transformation gives a flow, called an \textbf{RG flow}, in the
space of all permissible actions.
\begin{figure}
\begin{center}
\epsfig{file=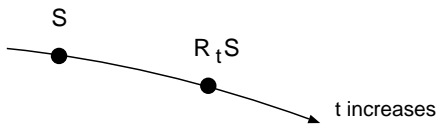}
\caption{The RG transformation $R_t$ by scale $\e^t$ gives a flow in
  the space of all possible actions.}
\end{center}
\end{figure}
Along an RG flow, the correlation length changes trivially as
\[
\xi (R_t S) = \e^{- t} \xi (S)
\]
as a consequence of rescaling of space.  This simple equation implies
the following:
\begin{itemize}
\item If a theory is critical with $\xi = \infty$, it remains critical
  under RG.
\item If a theory is non-critical, its correlation length becomes less
  and less along its RG flow.
\end{itemize}

\subsection{Fixed points and relevant \& irrelevant parameters}

A \textbf{fixed point} is an action $S^*$ which is invariant under the
RG transformation.  At the fixed point we find
\[
\vev{\phi (\vec{r}_1) \cdots \phi (\vec{r}_n)}_{S^*}
 = (1 - n \gamma^* \Delta t) \vev{\phi (\vec{r}_1 \e^{- \Delta t}) \cdots
 \phi (\vec{r}_n \e^{- \Delta t})}_{S^*}
\]
where
\[
\gamma^* \equiv \gamma (S^*)
\]
Especially for $n=2$, we obtain\footnote{We can extend this equation
  to define the correlation function for arbitrary $r$.}
\[
\vev{\phi (\vec{r}) \phi (- \vec{r})}_{S^*} =
\frac{\mathrm{const}}{r^{2 \gamma^*}}\quad (r \gg 1)
\]
This implies that the correlation length is infinite:
\[
\xi \big|_{S^*} = + \infty
\]
The scale transformation is compensated by renormalization of the
field, and the physics at the fixed point $S^*$ is scale invariant
with no characteristic length scale.

The scale dimension of the scalar field, $\gamma^*$, is not the only
important quantity that characterizes the fixed point.  To explain
this, let us consider the RG flow in a small (but finite) neighborhood
of $S^*$, where we expect to be able to linearize the RG
transformation.  Let
\[
S [\phi] = S^* [\phi] + \delta S [\phi]
\]
The RG transformation is given by
\[
\partial_t \delta S [\phi] = L \cdot \delta S [\phi]
\]
where $L$ is a linear transformation.  We introduce eigenvectors of
$L$ by\footnote{We assume all the eigenvalues are real.}
\[
L v_i [\phi] = y_i v_i [\phi]\quad (y_1 \ge y_2 \ge \cdots)
\]
Using $v_i [\phi]$ we can expand the difference $\delta S [\phi]$:
\[
\delta S [\phi] = \sum_{i=1}^\infty g_i (t) v_i [\phi]
\]
Since the RG transformation is linearized in this basis, we obtain
\[
\frac{d}{dt} g_i = c_i g_i \Longrightarrow g_i (t) \propto \e^{y_i t}
\]
We can use the parameters $\{ g_i \}$ as the coordinates of the theory
space in the neighborhood of $S^*$.  We call $g_i$ with positive
eigenvalues $y_i > 0$ \textbf{relevant}, and those with negative
eigenvalues \textbf{irrelevant}.\footnote{Those with zero eigenvalues
are called \textbf{marginal}.  We assume the absence of marginal
parameters to simplify the discussion below.}  Like the scale
dimension $\gamma^*$ of the scalar field, the scale dimensions of the
relevant and irrelevant parameters characterize the fixed point $S^*$.

\subsection{Critical subspace and renormalized trajectory}

As an example, let us consider a fixed point $S^*$ with one relevant
parameter $g_1$ with scale dimension $y_1 > 0$.  All the other (an
infinite number of them) parameters are irrelevant.  Let $g_2$ be the
least irrelevant, meaning that its scale dimension $y_2 < 0$ is the
largest.  In an infinitesimal neighborhood of $S^*$, it is a good
approximation to keep just $g_1$ and $g_2$ satisfying the RG equations
\[
\left\lbrace\begin{array}{c@{~=~}l}
\frac{d}{dt} g_1 & y_1\, g_1\\
\frac{d}{dt} g_2 & y_2\, g_2
\end{array}\right.
\]
The solution is easily obtained as
\[
g_1 (t) = g_1 (0) \e^{y_1 t},\quad g_2 (t) = g_2 (0) \e^{y_2 t}
\]
As $t$ grows, $g_2$ goes to $0$ while $g_1$ grows unless $g_1= 0$ to
begin with.  All the other irrelevant parameters $g_{i \ge 3}$
approach zero faster than $g_2$.
\begin{figure}
\begin{center}
\epsfig{file=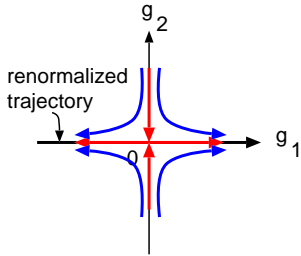}
\caption{As $t$ grows, the irrelevant parameter $g_2$ decreases.  The
  renormalized trajectory is the line $g_2=0$.}
\end{center}
\end{figure}
In the neighborhood of $S^*$, where $g_1$ is well defined, we can
define a subspace by a single condition
\[
\ffbox{g_1 = 0}
\]
This is called a \textbf{critical subspace} $\mathcal{S}_{cr}$.  The
critical subspace can be defined in a larger neighborhood of $S^*$ as
the subspace which flows into the fixed point $S^*$ as $t \to +
\infty$:
\[
\lim_{t \to \infty} R_t \mathcal{S}_{cr} = S^*
\]
This implies that the long-distance behavior of any theory in
$\mathcal{S}_{cr}$ is the same as the fixed point $S^*$.  Especially,
for any theory on $\mathcal{S}_{cr}$, we find the correlation length
infinite
\[
\xi \big|_{\mathrm{on~}\mathcal{S}_{cr}} = + \infty
\]
and the theory is critical.  Given an arbitrary theory with $N$
parameters, what is the likelihood that the theory is critical?  Since
$\mathcal{S}_{cr}$ is defined by a single condition $g_1 = 0$, we
expect that we need to fine tune only one of the $N$ parameters to
make the theory critical.

Another subspace, called a \textbf{renormalized trajectory}
$\mathcal{S}_\infty$, is defined by the vanishing of all the
irrelevant parameters:
\[
\ffbox{g_2 = g_3 = \cdots = 0}
\]
This is a one-dimensional subspace with a single parameter $g_1$.
Alternatively, $\mathcal{S}_\infty$ can be defined as the solution of
the equation
\[
(\forall t > 0)\quad R_t \mathcal{S}_\infty = \mathcal{S}_\infty
\]
implying that if $g_1$ is the coordinate of an arbitrary theory on
$\mathcal{S}_\infty$, then $g_1 \e^{- y_1 t}$ also belongs to
$\mathcal{S}_\infty$.  $\mathcal{S}_\infty$ can be also defined as the
RG flow coming out of the fixed point $S^*$.  Note that along the
renormalized trajectory, the correlation length is given by
\[
\xi = c_\pm |g_1|^{- \frac{1}{y_1}}
\]
where $c_\pm$ is a constant, and $\pm$ is for $\pm g_1 > 0$.
$\mathcal{S}_\infty$ is a one-dimensional subspace in the infinite
dimensional space.  Hence, a theory with a finite number of parameters
has \textbf{no chance} of lying in $\mathcal{S}_\infty$.

\subsection{Example: $3$ dimensional $\phi^4$ theory}

Let us now consider the $3$ dimensional $\phi^4$ theory defined by the
action
\[
S = \int_{p < 1} \frac{1}{2} \phi (\vec{p}) \phi (- \vec{p}) \left(
p^2 + m_0^2 \right) + \frac{\lambda_0}{4!} \int_{p_i < 1\atop p_1 +
\cdots + p_4 = 0} \phi ( \vec{p}_1) \cdots \phi ( \vec{p}_4)
\] 
For simplicity, we assume that the theory lies in the subspace where
we can linearize the RG transformation.  Then, the coordinates $g_i$
are obtained as functions of $\lambda_0$ and $m_0^2$:
\[
g_i = G_i (m_0^2, \lambda_0)\quad (i=1,2,\cdots)
\]
For any $\lambda_0 > 0$, we only need to tune $m_0^2 = m_{0, cr}^2
(\lambda_0)$ to make the theory critical.  This implies that there is
only one relevant parameter $g_1$, and $g_{i \ge 2}$ are all
irrelevant.  Hence, we obtain
\[
G_1 (m^2_{0, cr} (\lambda_0), \lambda_0) = 0
\]
The critical exponents $y_E, \eta$ introduced in the previous lecture
are given by the scale dimensions $y_1, \gamma^*$ of the fixed point
$S^*$:
\[
\left\lbrace\begin{array}{c@{~=~}l}
 y_E & y_1\\
 \eta & 2 \gamma^* - 1
\end{array}\right.
\]

For a given $\lambda_0$, let us consider a theory near criticality:
\[
m_0^2 = m_{0,cr}^2 (\lambda_0) + \Delta m_0^2
\]
Expanding $G_i$ with respect to $\Delta m_0^2$, we obtain
\[
\begin{array}{c@{~=~}l}
 g_1 & \Delta m_0^2 \frac{\partial G_1 (m_0^2, \lambda_0)}{\partial
 m_0^2} \Big|_{m_0^2 = m_{0,cr}^2} + \cdots\\
 g_{i \ge 2} & G_i (m_{0,cr}^2 (\lambda_0), \lambda_0) + \mathrm{O} \left(
 \Delta m_0^2 \right)
\end{array}
\]
Applying $R_t$ for large $t \gg 1$, we obtain
\[
\begin{array}{c@{~\simeq~}l}
g_1 (t) & \e^{y_1 t} \Delta m_0^2 \frac{1}{z_m (\lambda_0)}\\
g_{i \ge 2} (t) & \mathrm{O} (\e^{y_i t}) 
\end{array}
\]
where
\[
\frac{1}{z_m (\lambda_0)} \equiv \left.
\frac{\partial G_1 (m_0^2,
 \lambda_0)}{\partial m_0^2} \right|_{m_0^2 = m_{0,cr}^2 (\lambda_0)}
\]
Hence, by choosing
\[
\Delta m_0^2 = z_m (\lambda_0) \e^{- y_1 t} c_1
\]
for $t \gg 1$, we obtain
\[
\left\lbrace
\begin{array}{c@{~\simeq~}l}
 g_1 (t) & c_1\\
 g_{i \ge 2} (t) & 0
\end{array}\right.
\]
Thus, RG gives
\begin{eqnarray*}
&&\vev{\phi (\vec{r}_1 \e^{t}) \cdots \phi (\vec{r}_n
  \e^{t})}_{m_0^2 = m_{0,cr}^2 (\lambda_0) + \Delta m_0^2,
  \lambda_0}\\
&=& \exp \left[ - n \int_0^t dt' \gamma (R_{t'} S) \right]
 \vev{\phi (\vec{r}_1) \cdots \phi (\vec{r}_n)}_{g_1 = c_1,~g_{i \ge
  2} = 0}
\end{eqnarray*}
Defining
\[
z (\lambda_0) \equiv \exp \left[ 2 \int_0^\infty dt \left\lbrace
  \gamma \left(R_t S (m_{0,cr}^2, \lambda_0) \right) - \gamma^*
  \right\rbrace \right]
\]
we obtain
\[
 \vev{\phi (\vec{r}_1) \cdots \phi (\vec{r}_n)}_{g_1 = c_1,~g_{i \ge
  2} = 0} = z (\lambda_0)^{\frac{n}{2}} \lim_{t \to \infty} \e^{n
  \gamma^* t} \vev{\phi (\vec{r}_1 \e^{t}) \cdots \phi (\vec{r}_n
  \e^{t})}_{m_0^2, \lambda_0}
\]
where
\[
m_0^2 = m_{0, cr}^2 (\lambda_0) + z_m (\lambda_0) \e^{- y_1 t} c_1
\]
This reproduces the prescription introduced in the previous lecture.
\begin{figure}
\begin{center}
\epsfig{file=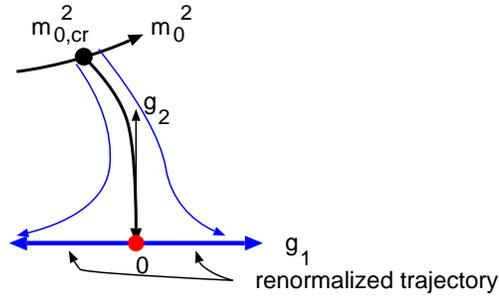}
\caption{The RG flows starting from near the critical point $m_{0,
    cr}^2 (\lambda_0)$ approach the renormalized trajectories.}
\end{center}
\end{figure}

We now understand that the renormalized trajectory
$\mathcal{S}_\infty$ corresponds to the renormalized theory viewed at
different scales.  As we have remarked already, it is practically
impossible to construct an action on $\mathcal{S}_\infty$.  However,
the correlation functions on $\mathcal{S}_\infty$ can be obtained as
the continuum limit of a theory only with a finite number of
parameters such as $m_0^2$ and $\lambda_0$.

\subsection{Three issues with renormalization}

Let us now examine the three issues with renormalization with which we
started this lecture.
\begin{enumerate}
\item \textbf{how to renormalize a theory} --- we now understand why
  the prescription given in the first lecture gives the continuum
  limit.
\item \textbf{universality} --- the renormalized theory is determined
  by the renormalized trajectory.  Hence, it is independent of
  $\lambda_0$.  We can go further.  Even if we start from an action
  with other interactions such as the $\phi^6$ interaction, as long as
  the critical theory lies on the same $\mathcal{S}_{cr}$, we get the
  same continuum limit.  As an extreme case, we get the same limit
  from the Ising model. 
\item \textbf{finite number of parameters} --- the number of
  renormalized parameters is determined by the dimensionality of
  $\mathcal{S}_\infty$, which is the number of relevant parameters.
\end{enumerate}
Thus, we understand that for each fixed point $S^*$ and the associated
renormalized trajectory $\mathcal{S}_\infty$, we can construct a
renormalized theory.
\begin{figure}[b]
\begin{center}
\epsfig{file=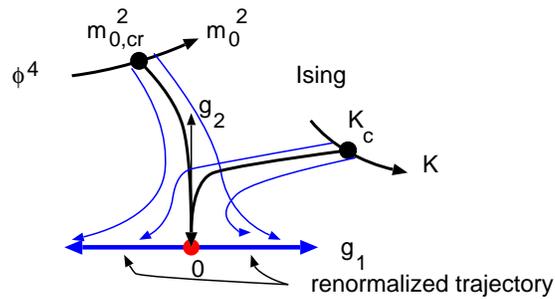}
\caption{As long as the critical point belongs to the same critical
  subspace $\mathcal{S}_{cr}$, we get the same continuum limit.}
\end{center}
\end{figure}

\section{Lecture 3 -- Perturbative Exact RG}

It was Joe Polchinski who first applied Wilson's RG to the
perturbative $\phi^4$ theory in $4$ dimensions in order to
``simplify'' the proof of renormalizability.\cite{Polchinski:1983gv}
His proof was simple in the sense that it did not rely on the analysis
of Feynman graphs or Feynman integrals.  I quoted the word ``simple'',
though, because even his proof is not what I would recommend first
year grad students to go through.\footnote{But the reading of his
summary of the idea of Wilson's RG is strongly recommended.}

We recall that Wilson's RG consists of three steps: integration of
momenta near the cutoff, rescaling of space, and renormalization of
field.  The second and third steps are crucial for the RG
transformation to have fixed points.  For applications to perturbation
theory, however, only the first step is important, since there is no
non-trivial fixed point in perturbation theory.  Hence, we define the
exact RG only by the first step, namely the lowering of the momentum
cutoff.

We consider a real scalar theory in $4$ dimensions in this lecture,
and QED in the next lecture.  We define the scalar theory by the
following action:
\begin{eqnarray*}
&&S (t) = \left. \frac{1}{2} \int_{p} \phi (p) \phi (-p)
\frac{p^2 + m^2}{K \left( \frac{p}{\mu \e^{-t}} \right)} \right\rbrace
= S_{free} (t) \\ && ~\left. - \sum_{n=1}^\infty \frac{1}{(2n)!}
\int_{p_1 + \cdots + p_{2n} = 0} \V_{2n} (t;
p_1, \cdots, p_{2n}) \phi (p_1) \cdots \phi (p_{2n}) \right\rbrace =
S_{int} (t)
\end{eqnarray*}
where we have omitted the tilde from the Fourier modes.  The function
$K(x)$ has the property that it is $1$ for $0 \le x \le 1$ and
converges rapidly to $0$ as $x \to \infty$.
\begin{figure}
\begin{center}
\epsfig{file=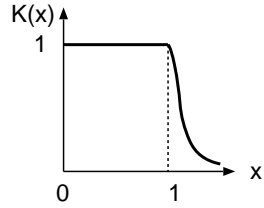}
\caption{The cutoff function $K (x)$ decreases rapidly in the region
  $x > 1$.}
\end{center}
\end{figure}
Hence, effectively the scalar field has the momentum cutoff
\[
\ffbox{
\Lambda = \mu \e^{-t}
}
\]
Please note that we have chosen a sign convention for $t$ so that
$\Lambda$ decreases as we increase $t$.  The interaction vertices
$\V_{2n}$ depend on momenta, and we treat them perturbatively, using
\[
\ffbox{
\frac{K\left(\frac{p}{\mu \e^{-t}}\right) }{p^2 + m^2}
}
\]
as the propagator.  Denoting the interaction vertices as
\begin{center}
\epsfig{file=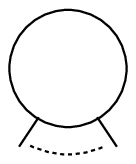}
\end{center}
a typical Feynman graph would look like
\begin{center}
\epsfig{file=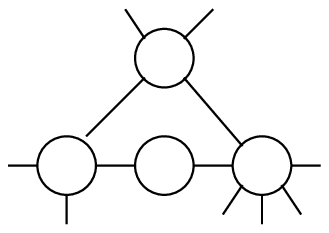}
\end{center}
Since the loop momenta are cut off at $p = \mu \e^{-t}$, there is
\textbf{no UV divergences}.

\subsection{Derivation of ERG differential equations}
				   
We define ERG by the change of the action under the decrease of
$\Lambda$ or equivalently the increase of $t$.  We change the action
in such a way that the correlation functions do not change.
\[
\vev{\phi (\vec{p}_1) \cdots \phi (\vec{p}_{2n})}_{S (t)}
= \vev{\phi (\vec{p}_1) \cdots \phi (\vec{p}_{2n})}_{S(t+\Delta t)}
\]
where $\Delta t > 0$ and
\[
(\forall i)\quad p_i < \mu \e^{- t - \Delta t}
\]
This can be done by compensating the change of the propagator by
changing the vertices, as we will see shortly.

In calculating the correlation functions, a propagator acts in one of
the three ways:
\begin{enumerate}
\item it connects two distinct vertices (internal line)
\item it connects the same vertex (internal line)
\item it is attached to an external source (external line)
\end{enumerate}
Now, as we increase $t$ infinitesimally to $t + \Delta t$, we find
\[
K \left(\frac{p}{\mu \e^{-t}}\right) = K \left(\frac{p}{\mu \e^{-
t-\Delta t}}\right) + \Delta t \underbrace{(-)
\frac{\partial}{\partial t} K \left(\frac{p}{\mu\e^{-t}}
\right)}_{\equiv \Delta \left(\frac{p}{\mu \e^{-t}}\right)}
\]
Hence, denoting
\begin{center}
\parbox{2cm}{\eq{\frac{\Delta \left( \frac{p}{\mu \e^{-t}}
      \right)}{p^2 + m^2} =}}
\parbox{2cm}{\epsfig{file=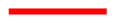}}
\end{center}
we obtain
\begin{center}
\parbox{5cm}{\eq{\frac{K\left( \frac{p}{\mu
	\e^{-t}}\right)}{p^2 + m^2} =
\frac{K \left(\frac{p}{\mu \e^{- t-\Delta t}}\right)}{p^2 +
	m^2} + \Delta t}}
\parbox{2cm}{\epsfig{file=3-redline.eps}}
\end{center}
Let us consider a Feynman diagram for the action $S(t)$.  The
propagator is given by the left-hand side above.  Imagine replacing
the propagator by the right-hand side and expand the diagram in powers
of $\Delta t$.  At first order in $\Delta t$, we must use the second
term (red line) just once.  It enters the Feynman graph in one of the
two ways shown in Figure 8,
\begin{figure}
\begin{center}
\epsfig{file=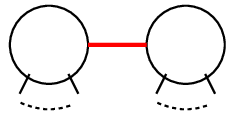} \hspace{2cm}\epsfig{file=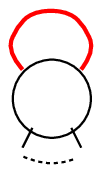}
\caption{Two types of corrections to $S_{int}$}
\end{center}
\end{figure}
We don't have to consider a red line attached to an external source,
if we restrict all the external momenta by $p < \mu \e^{- t - \Delta
  t}$.

Incorporating the contribution of the red line as part of a vertex, we
obtain the following diagrammatic differential equation:
\begin{center}
\parbox{4cm}{\eq{\frac{\partial}{\partial t} \V_{2n} (t; p_1, \cdots,
    p_{2n}) =}}
\parbox{2.5cm}{\epsfig{file=3-type1.eps}}
\parbox{0.5cm}{$+$}
\parbox{2cm}{\epsfig{file=3-type2.eps}}
\end{center}
The first term on the right is actually a sum over all possible
partitions of $2n$ external momenta into two parts.  This can be
expressed nicely using a more abstract notation as
\[
\partial_t (- S_{int}) (t) = \frac{1}{2} \int_{p} \frac{\Delta \left(
  \frac{p}{\mu \e^{-t}} \right)}{p^2 + m^2} \left[ \frac{\delta ( -
  S_{int}) (t)}{\delta \phi (p)} \frac{\delta (- S_{int}) (t)}{\delta
  \phi (-p)} + \frac{\delta^2 (- S_{int}) (t)}{\delta \phi (p) \delta
  \phi (-p)} \right]
\]
\textbf{Problem 3-1}: Check this result.\\ These differential
equations for the vertices, which we call perturbative ERG
differential equations, were derived by
Polchinski.\cite{Polchinski:1983gv}

\subsection{Renormalization}

As is the case with any differential equations, we must specify
initial conditions to solve the ERG differential equations.  To obtain
the renormalized trajectories or equivalently the continuum limit, we
need to take the theory to criticality, as we saw in the previous
lectures.  Thanks to universality, it does not matter what initial
conditions to choose as long as we can tune them for criticality.  For
example, we can choose the initial condition at $t = t_0$ in the
following form:
\[
\left\lbrace
\begin{array}{r@{~=~}l}
\V_2 (t_0; p, -p) & \mu^2 \e^{- 2 t_0} g (\lambda, t_0) + m^2 z_m
(\lambda, t_0) + p^2 z (\lambda, t_0)\\
\V_4 (t_0; p_1, \cdots, p_4) & - \lambda (1 + z_\lambda (\lambda,
t_0))\\
\V_{2n \ge 6} (t_0; p_1, \cdots, p_{2n}) & 0
\end{array}\right.
\]
We determine $g (\lambda, t_0)$ so that the theory is critical at $m^2
= 0$.  In perturbation theory, we determine the functions $g, z_m, z$,
and $z_\lambda$ as power series in $\lambda$ so that the vertices
$\{\V_{2n} (t)\}$, for any finite $t$, are well defined in the limit
$t_0 \to - \infty$.  This is how renormalizability was proved in
ref.~\cite{Polchinski:1983gv}.
\begin{figure}
\begin{center}
\epsfig{file=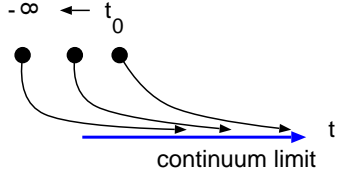}
\caption{The continuum limit is obtained as $t_0 \to - \infty$.}
\end{center}
\end{figure}

\subsection{Asymptotic conditions}

In the previous lecture we have mentioned that it is practically
impossible to construct the actions on the renormalized trajectories,
even though we can construct the continuum limit of the correlation
functions.  In perturbation theory, however, this can be done by
solving the ERG differential equations under appropriate conditions.

The $\phi^4$ theory does not have the kind of fixed point that the
three dimensional sibling has.  Hence, we cannot demand that the
renormalized trajectories trace back to a fixed point as we go
backward along ERG.  Instead, we can demand that the renormalized
trajectories satisfy the following \textbf{asymptotic conditions} as
$t \to - \infty$:\cite{Sonoda:2002pb}
\[
\left\lbrace\begin{array}{r@{~\longrightarrow~}l} \V_2 (t; p, - p) &
 \mu^2 \e^{- 2 t} a_2 (t; \lambda) + m^2 b_2 (t; \lambda)
 + p^2 c_2 (t; \lambda) \\ \V_4
 (t; p_1, \cdots, p_4) & a_4 (t; \lambda)\\
 \V_{2n \ge 6} (t; p_1, \cdots, p_{2n}) & \mathrm{O} (\e^{y_{2n} t})
\end{array}\right.
\]
Here, $a_2, b_2, c_2$, and $a_4$ are all power series of $\lambda$,
and at each order of $\lambda$ they are polynomials of $t$.  The
$t$-independent part of $b_2, c_2, a_4$ are not fixed by ERG, and we
can adopt the following convention:
\[
\ffbox{
\left\lbrace\begin{array}{l}
b_2 (0; \lambda) = c_2 (0; \lambda) = 0\\
a_4 (0; \lambda) = - \lambda
\end{array}\right.
}
\]
Hence, at first order in $\lambda$ we obtain
\[
\V_4^{(1)} = - \lambda
\]
This plays the role of a seed for perturbative expansions.  The
  vertices are uniquely determined with $m^2$ and $\lambda$ as
  parameters.\footnote{This was proved in ref.~\cite{Sonoda:2002pb}.}

As an example, let us determine the $2$-point vertex at first order.
It satisfies
\[
\partial_t \V_2^{(1)} (t; p, -p) = \frac{1}{2} \int_q \frac{\Delta
  \left( \frac{q}{\mu \e^{-t}} \right)}{q^2 + m^2}
  \underbrace{\V_4^{(1)} (t; q, -q, p, -p)}_{= - \lambda}
 = - \frac{\lambda}{2} \int_q \frac{\Delta
  \left( \frac{q}{\mu \e^{-t}} \right)}{q^2 + m^2}
\]
For large $-t \gg 1$ we obtain
\[
\int_q \frac{\Delta
  \left( \frac{q}{\mu \e^{-t}} \right)}{q^2 + m^2}
= \mu^2 \e^{- 2t} \int_q \frac{\Delta (q)}{q^2 + \frac{m^2}{\mu^2 \e^{- 2t}}}
\simeq \mu^2 \e^{-2t} \int_q \frac{\Delta (q)}{q^2} - m^2 \int_q
\frac{\Delta (q)}{q^4}
\]
Hence, we obtain
\begin{eqnarray*}
\V_2^{(1)} (t) &=& \frac{- \lambda}{2} \Bigg[ \int_{-\infty}^t dt'
 \int_q \Delta \left(\frac{q}{\mu \e^{-t'}}\right) \underbrace{\left(
 \frac{1}{q^2 + m^2} - \frac{1}{q^2} + \frac{m^2}{q^4} \right)}_{=
 \frac{m^4}{q^4 (q^2 + m^2)}}\\ && \quad - \frac{\mu^2 \e^{- 2t}}{2} \int_q
 \frac{\Delta (q)}{q^2} - t m^2 \int_q \frac{\Delta (q)}{q^4}\Bigg]\\
 &=& \frac{- \lambda}{2} \Bigg[ \int_q \left( 1 - K \left(
 \frac{q}{\mu \e^{-t}}\right) \right)\frac{m^4}{q^4
 (q^2 + m^2)} \\
&& \quad\quad\quad - \frac{\mu^2 \e^{- 2t}}{2} \int_q \frac{\Delta
 (q)}{q^2} - t m^2 \int_q \frac{\Delta (q)}{q^4} \Bigg]
\end{eqnarray*}
This implies
\[
\left\lbrace\begin{array}{c@{~=~}l} a_2^{(1)} (t) & \frac{\lambda}{4}
 \int_q \frac{\Delta (q)}{q^2}\\ b_2^{(1)} (t) & \frac{\lambda}{2} t
 \int_q \frac{\Delta (q)}{q^4} = \lambda t \frac{1}{(4 \pi)^2} 
\end{array}\right.
\]  
\textbf{Problem 3-2}: Compute $\V_2^{(1)}(t)$ explicitly for the
choice $K(x) = \theta (1-x)$.\\ \textbf{Problem 3-3}: Using
$\V_2^{(1)} (t)$ obtained above, compute the self-energy correction at
first order.  (Answer:
\[
\frac{- \lambda}{2} \int_{q < \mu \e^{-t}} \frac{1}{q^2 + m^2} +
\V_2^{(1)} (t) = \frac{- \lambda}{(4 \pi)^2} m^2 \ln \frac{m^2}{\mu^2}\quad)
\]

\subsection{Diagrammatic rules}

The vertices $\V_{2n} (t; p_1, \cdots, p_{2n})$ are most easily
calculated diagrammatically.  Consider a Feynman diagram with the
standard propagator, which we can decompose into the low and high
momentum parts:
\[
\frac{1}{p^2 + m^2} = \frac{K \left( \frac{p}{\mu \e^{-t}}
  \right)}{p^2 + m^2} + \frac{1 - K \left( \frac{p}{\mu \e^{-t}}
  \right)}{p^2 + m^2}
\]
Substituting this into each propagator of a Feynman diagram, we get a
bunch of diagrams in which low and high momentum propagators are
mixed.  By interpreting high momentum propagators not as propagators
but as part of vertices, we get Feynman diagrams only with low
momentum propagators.
\begin{figure}
\begin{center}
\epsfig{file=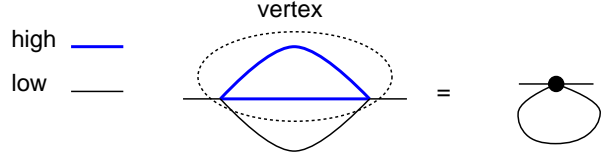}
\caption{We interpret high momentum propagators as part of vertices.}
\end{center}
\end{figure} 
Hence, the vertices are basically obtained from Feynman diagrams in
which the propagator is given by the high-momentum propagator.
The necessary UV subtractions are made in the BPHZ manner, i.e., at
the level of integrands we subtract the first couple of terms of the
Taylor series in $m^2$ and external momentum $p$.  We then add finite
counterterms to assure the correct $t$ dependence.\footnote{Details
  are still to be worked out.  See \cite{Sonoda:2005erg} for partial results.}

As an example, let us compute $\V_4$ at second order in $\lambda$.  In
the s-channel, we have 
\begin{center}
\parbox{3.5cm}{\epsfig{file=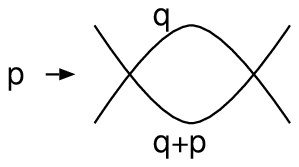}}
\parbox{7cm}{\eq{= \frac{\lambda^2}{2} \int_q \frac{\left(1 - K \left(
 \frac{q}{\mu \e^{-t}} \right) \right) \left( 1 - K \left(
 \frac{q+p}{\mu \e^{-t}} \right) \right)}{(q^2 + m^2)((q+p)^2 +
 m^2)}}}
\end{center}
As it is, this is UV divergent.  We subtract the integrand evaluated
at $p^2 = m^2=0$ to obtain a finite integral:
\[
\frac{\lambda^2}{2} \int_q \left[
\frac{\left(1 - K \left(
 \frac{q}{\mu \e^{-t}} \right) \right) \left( 1 - K \left(
 \frac{q+p}{\mu \e^{-t}} \right) \right)}{(q^2 + m^2)((q+p)^2 +
 m^2)} - \frac{\left( 1  - K \left(
 \frac{q}{\mu \e^{-t}} \right) \right)^2}{q^4} \right]
\]
But this does not have the correct $t$-dependence, and we must add a
finite counterterm:
\[
\lambda^2 t \, \int_q \frac{\Delta \left( \frac{q}{\mu \e^{-t}}
  \right) \left(1 - K \left( \frac{q}{\mu \e^{-t}} \right)
  \right)}{q^4}
= \lambda^2 t \, \int_q \frac{\Delta (q) (1 - K(q))}{q^4} = \lambda^2
  t \frac{1}{(4 \pi)^2}
\]
Another type of contribution to $\V_4$ comes from the following 1PR
(one-particle reducible) diagram\footnote{Note that the first term on
  the right-hand side of the ERG differential equation implies that the
  1PR diagrams also contribute to the vertices.}:
\begin{center}
\parbox{3cm}{\epsfig{file=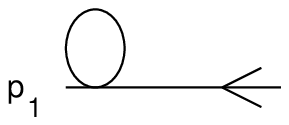}}
\parbox{5cm}{\eq{= \V_2^{(1)} (t) \frac{1 - \K{p_1}}{p_1^2 + m^2} ( -
    \lambda )}}
\end{center}

\subsection{Beta function and anomalous dimensions}

How do the beta function and anomalous dimensions arise in the context
of ERG?\footnote{For a detailed derivation of the results in this
subsection, please see the lecture notes \cite{Sonoda:2005erg}.}  To
understand this, we first note that besides $t$ the action $S (t)$ has
only $m^2, \lambda$, and $\mu$ as parameters.  The correlation
functions are independent of $t$, and hence we can use the notation
\[
\vev{\phi (\vec{p}_1) \cdots \phi (\vec{p}_{2n})}_{m^2, \lambda; \mu}
\]
This can be calculated using any action $S(t)$ as long as $(\forall
i)~\mu \e^{-t} > p_i$.

The beta function $\beta (\lambda)$ and anomalous dimensions $\beta_m
(\lambda), \gamma (\lambda)$ describe the $\mu$ dependence of the
correlation functions.  For $\Delta t$ infinitesimal, we find
\begin{eqnarray*}
&&\vev{\phi (\vec{p}_1) \cdots \phi (\vec{p}_{2n})}_{m^2 (1 - \Delta t
  \beta_m (\lambda)), \lambda - \Delta t \beta (\lambda); \mu (1 +
  \Delta t)}\\
&=& \left( 1 - \Delta t \cdot 2n \gamma (\lambda) \right)
 \vev{\phi (\vec{p}_1) \cdots \phi (\vec{p}_{2n})}_{m^2, \lambda; \mu}
\end{eqnarray*}
The beta function and anomalous dimensions are given in terms of
asymptotic vertices defined by the following behavior as $t \to -
\infty$:
\[
\left\lbrace\begin{array}{c@{~\longrightarrow~}l}
\V_4 (t; q \mu \e^{-t}, - q \mu \e^{-t}, p,-p) & a_4 (t; q) +
      \frac{m^2}{\mu^2 \e^{- 2t}} b_4 (t; q) + \frac{p^2}{\mu^2 \e^{-
      2t}} c_4 (t; q) \\ 
\V_6 (t; q \mu \e^{-t}, - q \mu \e^{-t},0,0,0,0) & \frac{1}{\mu^2
      \e^{- 2t}} a_6 (t; q)
\end{array}\right.
\]
Then, we obtain
\begin{eqnarray*}
2 \gamma (\lambda) &=& \frac{ - \frac{1}{2} \int_q \frac{\Delta
    (q)}{q^2} c_4 (0; q)}{1 - \frac{1}{2} \int_q
    \frac{K(q)(1-K(q))}{q^4} c_4 (0;q)}\\ 
\beta_m (\lambda) &=&
    \frac{1}{1 - \frac{1}{2} \int_q \frac{K(q)(1-K(q))}{q^2} a_4
    (0;q)} \Bigg[ \frac{1}{2} \int_q \Delta (q) \left( \frac{a_4
    (0;q)}{q^4} - \frac{b_4 (0;q)}{q^2}\right) \\
&& \quad - 2 \gamma (\lambda)
    \left\lbrace 1 + \frac{1}{2} \int_q K(q)(1-K(q)) \left( \frac{a_4
    (0;q)}{q^4} - \frac{b_4 (0;q)}{q^2} \right) \right\rbrace \Bigg]\\
\beta (\lambda) &=& - \frac{1}{2} \int_q \frac{\Delta (q)}{q^2} a_6 (0;q) -
    2 \gamma (\lambda) \left( 2 \lambda - \frac{1}{2} \int_q
    \frac{K(q)(1-K(q))}{q^2} a_6 (0;q)\right)
\end{eqnarray*}
These get much simplified for the particular choice:\footnote{For this
  choice it can be shown that the asymptotic vertices $b_2 (t;
  \lambda), c_2 (t; \lambda)$, and $a_4 (t; \lambda)$ are determined
  by $\beta, \beta_m, \gamma$.  For example,
\[
a_4 (t; \lambda) = \exp \left[ 4 \int_0^t dt'\, \gamma (\bar{\lambda}
  (t'; \lambda) ) \right] \cdot ( - \bar{\lambda} (t; \lambda) )
\]
 where $\bar{\lambda} (t; \lambda)$ is the running coupling satisfying
 the initial condition $\bar{\lambda} (0;\lambda) = \lambda$.}
\[
K (q) = \theta (1 - q)
\]
\textbf{Problem 3-4}: Simplify the formulas for the beta function and
anomalous dimensions for the above choice of $K$.  (Note $K(1-K) =
0$.)

To lowest order in $\lambda$, we find
\[
\left\lbrace\begin{array}{c@{~=~}l}
 a_4 (0;q) & - \lambda + \mathrm{O} (\lambda^2)\\
 b_4 (0;q) & \mathrm{O} (\lambda^2)\\
 c_4 (0;q) & \mathrm{O} (\lambda^2)\\
 a_6 (0;q) & 6 \lambda^2 \frac{1 - K(q)}{q^2} + \mathrm{O} (\lambda^3)
\end{array}\right.
\]
and we obtain the familiar results:\footnote{Our convention for $\beta
  (\lambda)$ differs from the standard one by the sign.}
\[
\left\lbrace\begin{array}{c@{~=~}l}
 \gamma (\lambda) & \mathrm{O} (\lambda^2)\\
 \beta_m (\lambda) & \frac{1}{2} \int_q \frac{\Delta (q)}{q^4} (-
 \lambda) + \mathrm{O} (\lambda^2) \simeq - \frac{\lambda}{(4
 \pi)^2}\\
 \beta (\lambda) & - \frac{1}{2} \int_q \frac{\Delta (q)}{q^2} 6
 \lambda^2 \frac{1 - K(q)}{q^2} + \mathrm{O} (\lambda^3) \simeq -
 \frac{3 \lambda^2}{(4 \pi)^2}
\end{array}\right.
\]

\section{Lecture 4 -- Application to QED}

To define QED we must introduce a vector field $A_\mu$ for photons and
a spinor field $\psi$ for electrons.  The free part of the action is
given by
\[
S_{free} (t) =  \frac{1}{2} \int_k A_\mu (k) A_\nu (-k) \frac{k^2
\delta_{\mu\nu} - \left(1 - \frac{1}{\xi} \right) k_\mu k_\nu}{K
  \left( \frac{k}{\mu \e^{-t}} \right)}
 + \int_p \bar{\psi} (-p) \frac{\fmslash{p} + i m}{\K{p}} \psi
(p)
\]
so that the propagators are given by
\begin{center}
\parbox{3cm}{\epsfig{file=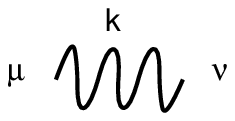}}\parbox{6cm}{\eq{=
    \frac{\K{k}}{k^2} \left( \delta_{\mu\nu} - (1-\xi) \frac{k_\mu
    k_\nu}{k^2} \right)}}\\
\parbox{2cm}{\epsfig{file=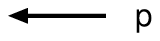}}\parbox{4cm}{\eq{=
    \frac{\K{p}}{\fmslash{p} + i m}}}
\end{center}

We write the interaction part of the action as
\begin{eqnarray*}
&&S_{int} (t) = - \sum_{M=1}^\infty \sum_{N=1}^\infty \frac{1}{M!}
\frac{1}{(N!)^2} \int_{k_1 + \cdots + p_N = 0} \\
&& ~A_{\mu_1} (k_1) \cdots
A_{\mu_M} (k_M) \bar{\psi} (- q_1) \cdots \bar{\psi} (- q_N)\\
&& ~\cdot \V_{\mu_1 \cdots \mu_M, N} (-t; k_1, \cdots, k_M; - q_1,
\cdots, - q_N; p_1, \cdots, p_N) \psi (p_1) \cdots \psi (p_N)
\end{eqnarray*} 
We can denote the vertices graphically as
\begin{center}
\parbox{7cm}{\eq{\V_{\mu_1,\cdots,\mu_M,N} (t; k_1, \cdots, - q_1,
    \cdots, p_1, \cdots ) = }}
\parbox{3cm}{\epsfig{file=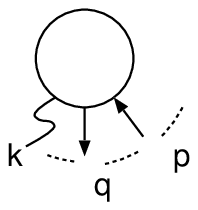}}
\end{center}
The vertices satisfy the ERG differential equations given graphically
by
\begin{center}
\parbox{1cm}{\eq{\frac{\partial}{\partial t}}}
\parbox{2cm}{\epsfig{file=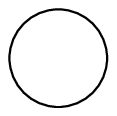}}
\parbox{1cm}{\eq{=}}
\parbox{3cm}{\epsfig{file=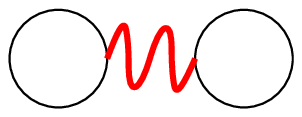}}\\
\parbox{1cm}{\eq{+}}
\parbox{3cm}{\epsfig{file=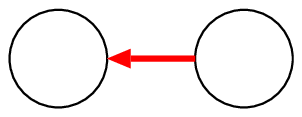}}
\parbox{1cm}{\eq{+}}
\parbox{2cm}{\epsfig{file=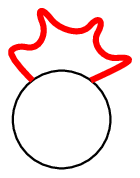}}
\parbox{1cm}{\eq{+}}
\parbox{2cm}{\epsfig{file=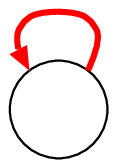}}
\end{center}
where
\begin{center}
\parbox{3cm}{\epsfig{file=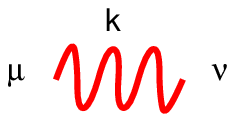}}
\parbox{6cm}{\eq{= \frac{\D{k}}{k^2} \left( \delta_{\mu\nu} - (1-\xi)
    \frac{k_\mu k_\nu}{k^2}\right)}}\\
\parbox{2cm}{\epsfig{file=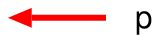}}
\parbox{4cm}{\eq{= \frac{\D{p}}{\fmslash{p} + i m}}}
\end{center}
The theory is obtained by solving the ERG differential equations under
the following asymptotic conditions for $t \to - \infty$:
\begin{eqnarray*}
\V_{\mu\nu} (t; k, -k) &\longrightarrow& \delta_{\mu\nu} \left( \mu^2
\e^{- 2 t} a_2 (t) + m^2 b_2 (t) \right)\\
&& ~+ k^2 \delta_{\mu\nu} c_2 (t) + k_\mu k_\nu d_2 (t)\\
\V_1 (t; p, -p) &\longrightarrow& a_f (t) \fmslash{p} + b_f (t) i m\\
\V_{\mu,1} (t; k, -p-k,p) &\longrightarrow& a_3 (t) \gamma_\mu\\
\V_{\alpha\beta\gamma\delta} (t; k_1, \cdots, k_4) &\longrightarrow&
a_4 (t) \left( \delta_{\alpha\beta} \delta_{\gamma\delta} +
\delta_{\alpha\gamma} \delta_{\beta\delta} + \delta_{\alpha\delta}
\delta_{\beta\gamma} \right)\\
\mathrm{higher~vertices} &\longrightarrow& 0
\end{eqnarray*}
Note that the three-point vertex of photons vanish identically 
\[
\V_{\alpha\beta\gamma} = 0
\]
since we impose invariance under \textbf{charge conjugation}.

In specifying the asymptotic conditions, all the $t$-dependence of the
asymptotic vertices are fixed by the ERG differential equations.  But
\[
b_2 (0),~c_2 (0),~d_2 (0),~a_f (0),~b_f (0),~a_3 (0),~a_4 (0)
\]
are free parameters, and we can take them arbitrarily as far as
renormalization is concerned.  We can choose
\[
c_2 (0) = a_f (0) = 0
\]
by normalizing the fields $A_\mu, \psi$ appropriately.  We can also
take
\[
b_f (0) = 0
\]
by normalizing the mass parameter $m$ of the electron appropriately.
We could choose $a_3 (0) = e$, the electric charge, but we prefer to
introduce $e$ through the Ward identities.  In the following we will
show that 
\[
b_2 (0),~d_2 (0),~a_3 (0),~a_4 (0)
\]
are determined uniquely as power series of $e$ by imposing that the
theory satisfy the Ward identities.

\subsection{Ward identities}

Let us recall the Ward identities.  The two-point function of the
gauge field satisfies
\[
\frac{1}{\xi} k_\mu \vev{A_\mu (-k) A_\nu (k)} = \frac{k_\nu}{k^2}
\]
and those with electron fields satisfy\footnote{Here we only consider
  the connected part of the correlation.}
\begin{eqnarray*}
&&\frac{1}{\xi} k_\mu \vev{A_\mu (-k) A_{\mu_1} \cdots A_{\mu_M} (k_M)
\psi (q_1) \cdots \psi (q_N) \bar{\psi} (-p_1) \cdots \bar{\psi} (-
p_N)}\\
&=& \frac{e}{k^2} \sum_{n=1}^N \left( \vev{A_{\mu_1} \cdots \psi (q_n
    - k) \cdots} - \vev{A_{\mu_1} \cdots \bar{\psi} (- p_n - k)
    \cdots} \right)
\end{eqnarray*}

These Ward identities are expected to imply the invariance of the
action under gauge transformations or the BRST transformation.  The
Ward identities imply
\begin{enumerate}
\item that the photon has only two transverse degrees of freedom,
\item that the S-matrix is independent of the gauge fixing parameter $\xi$.
\end{enumerate}

There is a complication, however, since our action
\[
S (t) = S_{free} (t) + S_{int} (t)
\]
gives the correlation functions correctly only for external momenta
less than $\mu \e^{- t}$.  This calls for two changes:
\begin{enumerate}
\item We must modify the BRST transformation.
\item The action is not strictly invariant under the modified BRST
  transformation.
\end{enumerate}
The derivation would take too much space and time (one full lecture;
see ref.~\cite{Sonoda:2005erg}),
and we will content ourselves by merely describing the results.

\subsubsection{Modified BRST transformation}

We generalize the action by including Faddeev-Popov ghost and
antighost fields:
\begin{eqnarray*}
S_{free} (t) &=& \frac{1}{2} \int_k A_\mu (-k) A_\nu (k) \frac{k^2
  \delta_{\mu\nu} - \left( 1 - \frac{1}{\xi} \right) k_\mu
  k_\nu}{\K{k}}\\
&& \quad + \int_p \bar{\psi} (-p) \frac{\fmslash{p} + i m}{\K{p}} \psi
  (p) + \int_k \bar{c} (-k) \frac{k^2}{\K{k}} c (k)
\end{eqnarray*}
where $c$ and $\bar{c}$ are both anticommuting fields.  The
interaction part of the action is free of $c, \bar{c}$.  We then
introduce the following BRST transformation:
\[
\left\lbrace\begin{array}{c@{~=~}l}
 \delta_\ep A_\mu (k) &  k_\mu \ep~c (k) \\
 \delta_\ep c (k) & 0\\
 \delta_\ep \bar{c} (-k) & - \frac{1}{\xi} k_\mu A_\mu (-k) \ep\\
 \delta_\ep \psi (p) & e \int_k \ep~ c (k) \frac{\K{p}}{\K{p-k}} \psi (p-k)\\
 \delta_\ep \bar{\psi} (-p) & - e \int_k \ep~ c (k)
 \frac{\K{p}}{\K{p+k}} \bar{\psi} (-p-k)
\end{array}
\right.
\]
where $\ep$ is an arbitrary anticommuting constant.  In the limit $t
\to - \infty$, we find $K = 1$, and the above reduces to the standard
BRST transformation.

\subsubsection{BRST transformation of the action}

The total action $S(t) = S_{free} (t) + S_{int} (t)$ is not quite BRST
invariant.  It must transform as
\[
\delta_\ep S (t) = - \int_k \ep~ c (k) \mathcal{O} (t; -k)
\]
where
\begin{eqnarray*}
\mathcal{O} (t;-k) &\equiv& \int_p \Bigg[
 \frac{\overrightarrow{\delta}}{\delta \bar{\psi}_i (-p)} \left( - S
 (t) \right) \cdot \left( - S(t) \right)
 \frac{\overleftarrow{\delta}}{\delta \psi_j (p+k)}\\ &&+
 \frac{\overrightarrow{\delta}}{\delta \bar{\psi}_i (-p)} \left( - S
 (t) \right) \frac{\overleftarrow{\delta}}{\delta \psi_j (p+k)}
 \Bigg] U_{ji} (t; -p-k,p)
\end{eqnarray*}
and
\[
U (t; - p-k,p) \equiv e \left\lbrace \K{p+k} \frac{1 -
  \K{p}}{\fmslash{p} + i m} - \frac{1 - \K{p+k}}{\fmslash{p} +
  \fmslash{k} + i m} \K{p} \right\rbrace
\]
The above BRST transformation property guarantees the Ward identities
of the correlation functions for external momenta less than the cutoff
$\mu \e^{-t}$.  It also has the important property that if it is
satisfied at some $t$, the ERG differential equation implies that it
is satisfied at any other $t$.  Hence, the BRST transformation
property is consistent with the ERG.

\subsubsection{BRST invariance at $t \to - \infty$}

The consistency of the BRST ``invariance'' or transformation property
of the action $S(t)$ with ERG implies that we only need to check it
asymptotically as $t \to - \infty$.  If it is satisfied
asymptotically, it is satisfied for any finite $t$.  As $t \to
-\infty$,
\[
\delta_\ep S (t) = - \int_k \ep~ c (k) \mathcal{O} (t; -k)
\]
gives the following three equations.
\begin{center}
\parbox{9cm}{\eq{k_\mu \left( \mu^2 \e^{- 2 t} a_2 (t) + m^2 b_2 (t) +
    k^2 (c_2 + d_2) (t) \right) = - \lim_{t \to - \infty}}}
\parbox{3cm}{\epsfig{file=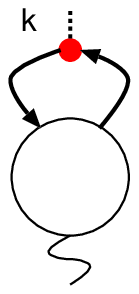}}\\
\parbox{5cm}{\eq{a_3 (t) \fmslash{k} = e \left( 1 - a_f (t) \right)
    \fmslash{k} - \lim_{t \to - \infty}}}
\parbox{3cm}{\epsfig{file=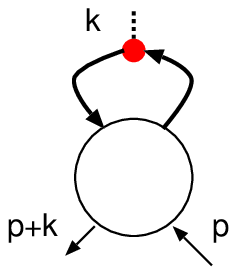}}\\
\parbox{7cm}{\eq{a_4 (t) \left( k_\alpha \delta_{\beta\gamma} +
    k_\beta \delta_{\gamma\alpha} + k_\gamma \delta_{\alpha\beta}
    \right) = - \lim_{t \to - \infty}}}
\parbox{3cm}{\epsfig{file=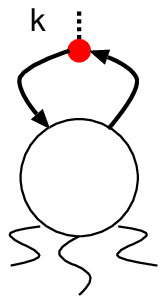}}
\end{center}
where
\begin{center}
\parbox{3cm}{\epsfig{file=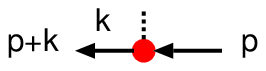}}
\parbox{3cm}{\eq{= U (t; -p-k,p)}}
\end{center}
For the higher point vertices, we simply get $0=0$.  From the second
equation, we immediately see
\[
\ffbox{
a_3 (0) = e + \mathrm{O} (e^3)
}
\]
It is clear that the above BRST transformation properties determine
the coefficients $b_2 (0), d_2 (0), a_3 (0), a_4 (0)$ uniquely if we
use the convention $c_2 (0) = a_f (0) = 0$.

\subsection{One-loop calculations}

Let us compute $b_2 (0), d_2 (0), a_3 (0) - e, a_4 (0)$ at one-loop.

\subsubsection{Photon two-point}

At one-loop, we obtain
\begin{center}
\parbox{0.5cm}{$-$}
\parbox{1.5cm}{\epsfig{file=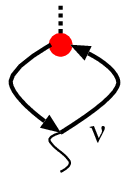, width=1.5cm}}
\parbox{8cm}{\eq{= e^2 \int_q }\hfill}\\
\parbox{12cm}{\hfill\eq{\times \Sp \left[ \gamma_\nu \left(
\K{q+k} \frac{1 - \K{q}}{\fmslash{q} + i m} - \frac{1 -
  \K{q+k}}{\fmslash{q} + \fmslash{k} + i m} \K{q} \right) \right]}}
\end{center}
\[
\stackrel{t \to \infty}{\longrightarrow}
 e^2 k_\nu \left[ - 2 \mu^2 \e^{-2t} \int_q \frac{\Delta (q) (1 -
 K(q))}{q^2} + \left( m^2 + \frac{1}{3} k^2\right) \int_q \frac{\Delta
 (q)}{q^4} \right]
\]
Thus, we obtain
\[
\left\lbrace
\begin{array}{c@{~=~}l}
 b_2 (0) & e^2 \int_q \frac{\Delta (q)}{q^4} = \frac{2 e^2}{(4
 \pi)^2}\\ c_2 (0) & \frac{e^2}{3} \int_q \frac{\Delta (q)}{q^4} =
 \frac{4 e^2}{3 (4 \pi)^2}
\end{array}\right.
\]
(In fact $b_2 (t)$ is independent of $t$.)

\subsubsection{Photon-electron vertex}

At one-loop, we obtain
\begin{center}
\parbox{0.5cm}{$-$}
\parbox{8cm}{\epsfig{file=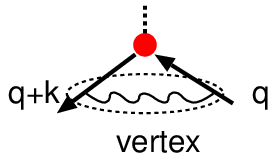, width=3cm}\hfill}
\end{center}
\begin{eqnarray*}
&=& - e^3 \int_q \frac{1 - \K{q-p}}{(q-p)^2} \,\gamma_\mu \Bigg[
 \K{q+k} \frac{1 - \K{q}}{\fmslash{q} + i m} \\ && - \frac{1 -
 \K{q+k}}{\fmslash{q} + \fmslash{k} + i m} \K{q} \Bigg] \gamma_\nu
 \left( \delta_{\mu\nu} - (1-\xi) \frac{(q-p)_\mu (q-p)_\nu}{(q-p)^2}
 \right)\\ &\stackrel{t \to \infty}{\longrightarrow}& - e^3
 \fmslash{k} \int_q \frac{1}{q^4} \left\lbrace \xi K (q) (1 - K(q))^2
 + \frac{3-\xi}{4} ( 1 - K(q) ) \Delta (q) \right\rbrace
\end{eqnarray*}
Hence, we obtain
\[
a_3 (0) = e \left[ 1 -  e^2 \left( \xi \int_q
 \frac{1}{q^4} K (q) (1 - K(q))^2 
+ \frac{3 - \xi}{4} \frac{1}{(4 \pi)^2}\right) \right]
\]

\subsubsection{Photon four-point}

At one-loop, we obtain
\begin{center}
\parbox{0.5cm}{$-$}
\parbox{3cm}{\epsfig{file=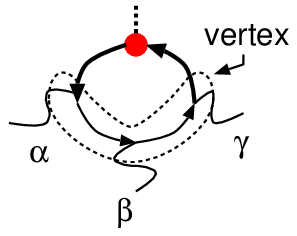, width=3cm}}
\parbox{5cm}{$-$ five permutations\hfill}
\end{center}
\begin{eqnarray*}
&=& e^4 \int_q \Sp \Bigg[ \gamma_\gamma \frac{1 -
      \K{q - k_3}}{\fmslash{q} - \fmslash{k}_3 + i m} \gamma_\beta 
    \frac{1 - \K{q - k_3 - k_2}}{\fmslash{q} - \fmslash{k}_3 -
      \fmslash{k}_2 + i m} \gamma_\alpha \\
&& \cdot \left\lbrace \K{q+k} \frac{1 - \K{q}}{\fmslash{q} + i m} -
      \frac{1 - \K{q+k}}{\fmslash{q} + \fmslash{k} + i m} \K{q}
      \right\rbrace \Bigg]\\
&& \qquad\qquad\qquad+~ \textrm{five permutations}\\
&\stackrel{t \to \infty}{\longrightarrow}& 2 e^4 
\left( k_\alpha \delta_{\beta\gamma} + k_\beta \delta_{\gamma\alpha} +
k_\gamma \delta_{\alpha\beta} \right) \underbrace{\int_q \frac{\Delta (q)
  (1 - K(q))^2}{q^4}}_{= \frac{2}{3 (4 \pi)^2}}
\end{eqnarray*}
Hence, $a_4 (t)$ is independent of $t$:
\[
a_4 (t) = \frac{4}{3 (4 \pi)^2} e^4
\]

\subsection{Comments on chiral QED}

We can construct chiral QED by replacing the electron field by
massless R-hand fermion fields $\psi_i$ with charge $e_i$.
\[
\gamma_5 \psi_i = \psi_i
\]
Unless there are equal numbers of positive and negative charged
particles, there is no invariance under charge conjugation, and the
three-photon vertex $\V_{\alpha\beta\gamma}$ is non-vanishing.  Its
asymptotic form is given by
\[
\V_{\alpha\beta\gamma} (t; k_1, k_2, k_3) \stackrel{t \to -
  \infty}{\longrightarrow} i \tilde{a}_{3} (t) \left( k_{1 \alpha}
  \delta_{\beta\gamma} + k_{2 \beta} \delta_{\gamma\alpha} + k_{3
  \gamma} \delta_{\alpha\beta} \right)
\]
Correspondingly, we get one more Ward identity for $t \to - \infty$:
\begin{center}
\parbox{7cm}{\eq{i \tilde{a}_3 (t) \left( k^2 \delta_{\alpha\beta} +
    k_{1 \alpha} k_{\beta} + k_\alpha k_{2 \beta} \right) = - \lim_{t
    \to \infty}}}
\parbox{3cm}{\epsfig{file=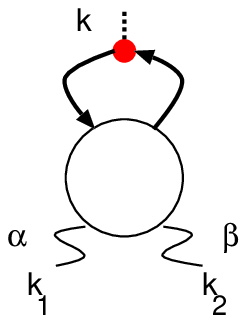}}
\end{center}
The potential problem is that the right-hand side gets an anomalous
contribution proportional to
\[
\epsilon_{\alpha\beta\mu\nu} k_{1 \mu} k_{2 \nu}
\]
Unless this coefficient vanishes automatically, we cannot satisfy the
Ward identities.

The non-renormalization theorem of anomaly is the statement that if
the coefficient of the above term vanishes at 1-loop level, i.e.,
\[
\sum_i e_i^3 = 0
\]
then it vanishes automatically at higher loop levels.  This has been
proved using the Callan-Symanzik equation (analogous to RG).  (See
\cite{Piguet:1980nr} for review.)  I am trying to prove the theorem
using the perturbative ERG method sketched in this lecture, hoping the
proof is simpler.

The dimensional regularization, which is the most popular method of
practical calculations, has trouble regularizing chiral gauge
theories.  I am also hoping that the ERG method will find practical
applications in constructing chiral non-abelian gauge theories
perturbatively.

\bibliography{pohang_06}

\newpage
\appendix

\section{$O(N)$ non-linear sigma model in $2$ dimensions}

The action is given by
\[
S = \frac{1}{2 g_0} \sum_{\vec{n}} \sum_{i=1,2} \left(
\Phi^I_{\vec{n}+\hat{i}} - \Phi^I_{\vec{n}} \right)^2
\]
where $g_0 > 0$, and $\Phi^I$ is an $N$-dimensional unit vector:
\[
\Phi^I \Phi^I = 1
\]
We have suppressed the summation over the repeated $I = 1,\cdots,N$.

The action is invariant under the continuous $O(N)$ transformation:
\[
\Phi^I_{\vec{n}} \longrightarrow O^I_{~J} \Phi^J_{\vec{n}}
\]
where $O^I_{~J}$ is an arbitrary $O(N)$ matrix independent of
$\vec{n}$.  In two dimensions this symmetry is never spontaneously
broken except at $g_0 = 0$.\footnote{In three and higher dimensions
  there is a critical point at a non-vanishing $g_0$.}

The continuum limit is obtained as
\[
\vev{\Phi^I (\vec{r}) \Phi^J (\vec{0})}_{g; \mu}
 \equiv \left( \frac{g}{1 + c~g} \right)^{2 \gamma}
 \lim_{t \to \infty} t^{2 \gamma} \vev{\Phi^I_{\vec{n} = \vec{r} \mu
 \e^t} \Phi^J_{\vec{0}}}_{g_0}
\]
where we choose $g_0$ as
\[
g_0 = \frac{1}{t + c \ln t - \ln \frac{\Lambda (g)}{\mu}}
\]
where 
\[
\frac{\Lambda (g)}{\mu} \equiv \e^{- \frac{1}{g}} \left( \frac{g}{1 +
  c~g} \right)^{- c}
\]
and
\[ 
\left\lbrace\begin{array}{c@{~\equiv~}l}
 c & \frac{1}{N-2}\\
 \gamma & \frac{1}{2} \frac{N-1}{N-2}
\end{array}\right.
\]

It is straightforward to derive the RG equation:
\[
\vev{\Phi^I (\vec{r} \e^{- \Delta t}) \Phi^J (\vec{0})}_{g + \Delta t
  \cdot \beta (g); \mu}
 = \left( 1 + \Delta t \cdot 2 \gamma~g \right) \vev{ \Phi^I (\vec{r})
  \Phi^J (\vec{0}) }_{g; \mu}
\]
where
\[
\beta (g) \equiv g^2 + c g^3
\]

The correlation length is of order $\frac{1}{\Lambda (g)}$, and for
short distances $r \ll \frac{1}{\Lambda (g)}$ the two-point function
can be expanded as
\[
\vev{\Phi^I (\vec{r}) \Phi^J (\vec{0})}_{g; \mu} = \delta^{IJ} \left(
\frac{g}{1 + c~g} \frac{1 + c~g (\ln \mu r )}{g (\ln \mu r )} \right)^{2
  \gamma} C (g (\ln \mu r))
\]
where $g(t)$ is the solution of
\[
\frac{d}{dt} g(t) = \beta (g(t))
\]
satisfying the initial condition $g(0) = g$, and $C (g)$ can be
expanded in powers of $g$.

\section{Large $N$ approximation for the lattice $\phi^4$ theory}

In this appendix we would like to compute the critical exponents $y_E,
\eta$ and follow the renormalization prescription given in lecture 1
explicitly for the $\phi^4$ theory in $3$ \& $4$ dimensions.  For this
purpose we generalize the $\phi^4$ theory by introducing $N$ scalar
fields.  For $N \gg 1$, we can compute $y_E, \eta$ using the mean
field approximation.

The action is given by
\[
S = \sum_{\vec{n}} \left( \frac{1}{2} \left( \phi^I_{\vec{n}+\hat{i}}
- \phi^I_{\vec{n}}\right)^2 + \frac{m_0^2}{2} (\phi^I_{\vec{n}})^2 +
\frac{\lambda_0}{8 N} \left(\left(\phi^I_{\vec{n}}\right)^2\right)^2
\right)
\]
where the summation symbol over $I=1,\cdots,N$ is suppressed.  We
expect the following symmetry breaking:
\begin{center}
\begin{tabular}{c|c}
\hline $m_0^2 < m_{0,cr}^2 (\lambda_0)$ & $m_0^2 > m_{0,cr}^2
(\lambda_0)$\\ \hline $\vev{\phi^I} = v \delta^{IN} \ne 0$ &
$\vev{\phi^I} = 0$\\ $O(N) \stackrel{\mathrm{broken}}{\to} O(N-1)$ &
$O(N)$ symmetric\\ $N-1$ massless Nambu-Goldstone particles& $N$
massive particles\\ \& $1$ massive particle & of the same mass
$m_{\mathrm{ph}}$\\ \hline
\end{tabular}
\end{center}

\subsection{$3$ dimensions}

\subsubsection{Introduction of an auxiliary field}

For large $N \gg 1$, only the average behavior of the fields $\phi^I$
becomes important, and this is the reason why the theory simplifies in
this limit.  Mathematically the large $N$ limit is equivalent to the
saddle point approximation of the integral
\begin{eqnarray*}
\int_{-\infty}^\infty dx\, \e^{- N f(x)} &=& \e^{- N f(x_0)}
 \sqrt{\frac{\pi}{N f''(x_0)}} \left( 1 + \mathrm{O}
 \left(\frac{1}{N}\right) \right)\\ &=& \frac{1}{\sqrt{N}} \exp
 \left[- N ( f(x_0) + \underbrace{\mathrm{O}
 (1/N)}_{\mathrm{corrections}} )\right]
\end{eqnarray*}
where $x_0$ is the minimum point of $f(x)$ so that 
\[
f'(x_0) = 0,\quad f''(x_0) > 0
\]

To take the large $N$ limit, it is convenient to introduce an
auxiliary real field $\alpha_{\vec{n}}$ to rewrite the action as
follows:
\begin{eqnarray*}
S &=& \sum_{\vec{n}} \Bigg[ \frac{1}{2} \left(
\phi^I_{\vec{n}+\hat{i}} - \phi^I_{\vec{n}}\right)^2 + \frac{m_0^2}{2}
(\phi^I_{\vec{n}})^2 + \frac{\lambda_0}{8 N}
\left((\phi^I_{\vec{n}})^2\right)^2\\ && \qquad + \frac{N}{2
\lambda_0} \left( \alpha_{\vec{n}} - \frac{i \lambda_0}{2 N}
(\phi^I_{\vec{n}})^2 \right)^2 \Bigg]
\end{eqnarray*}
The integration over $\alpha_{\vec{n}}$ simply changes the partition
function by a constant factor, but the correlation functions remain
intact.  Expanding the last term, we obtain
\[
S = \sum_{\vec{n}} \left( \frac{1}{2} \left( \phi^I_{\vec{n}+\hat{i}}
- \phi^I_{\vec{n}}\right)^2 + \frac{m_0^2}{2} (\phi^I_{\vec{n}})^2 +
\frac{N}{2 \lambda_0} \alpha_{\vec{n}}^2 - i \alpha_{\vec{n}}
\frac{1}{2} (\phi^I_{\vec{n}})^2 \right)
\]
without $\phi^4$ interaction terms.  

If we integrate over the $N$ fields $\phi^I_{\vec{n}}$ before
$\alpha_{\vec{n}}$, we get
\[
Z_1 [\alpha] \equiv \prod_{\vec{n}} \int d\phi_{\vec{n}} \exp \left[ -
 \frac{1}{2} \sum_{\vec{n}} \left( \sum_i (\phi_{\vec{n}+\hat{i}} -
 \phi_{\vec{n}})^2 + m_0^2 \phi_{\vec{n}}^2 - i \alpha_{\vec{n}}
 \frac{1}{2} \phi_{\vec{n}}^2 \right) \right]
\]
from each component $\phi^I$, and the partition function can be
written as
\[
Z = \prod_{\vec{n}} \int d\alpha_{\vec{n}} \, \e^{- \frac{N}{2
\lambda_0} \sum_{\vec{n}} \alpha_{\vec{n}}^2 } Z_1 [\alpha]^N =
\prod_{\vec{n}} \int d\alpha_{\vec{n}} \, \e^{- N \left( \frac{1}{2
\lambda_0} \sum_{\vec{n}} \alpha_{\vec{n}}^2 - \ln Z_1 [\alpha]
\right)}
\]
Thus, for $N \gg 1$, we can calculate the integrals over
$\alpha_{\vec{n}}$ using the saddle point approximation.  We decompose
\[
\alpha_{\vec{n}} = i \Delta m_0^2 + \sigma_{\vec{n}}
\]
where $i \Delta m_0^2$ is the value of $\alpha_{\vec{n}}$ at the
saddle point, and $\sigma_{\vec{n}}$ is the fluctuation around the
saddle point.  In the limit $N \to \infty$ we can ignore the
integration over the fluctuations $\sigma_{\vec{n}}$.

Now, the condition that $i \Delta m_0^2$ is a saddle point is given by
\[
\frac{\partial}{\partial \alpha_{\vec{n}}}
 \left( \frac{1}{2 \lambda_0} \sum_{\vec{n'}} \alpha_{\vec{n'}}^2 -
 \ln Z_1 [\alpha] \right)\Bigg|_{\alpha = i \Delta m_0^2}
 = 0
\]
Hence, we obtain
\[
\Delta m_0^2 - \vev{\frac{\lambda_0}{2} \phi_{\vec{n}}^2}_{S'} \equiv
 \prod_{\vec{n'}} \int d\phi_{\vec{n'}} \left(\Delta m_0^2 -
 \frac{\lambda_0}{2} \phi_{\vec{n}}^2\right) \e^{ - S' } = 0
\]
where 
\[
S' \equiv \sum_{\vec{n'}} \left( \frac{1}{2} \sum_i \left(
\phi_{\vec{n'}+\hat{i}} - \phi_{\vec{n'}} \right)^2 + \frac{m_0^2 +
\Delta m_0^2}{2} \phi_{\vec{n'}}^2 \right)
\]
This condition determines $\Delta m_0^2$.

From now on we assume the symmetric phase.  Let
\[
m_{0,\mathrm{ph}}^2 \equiv m_0^2 + \Delta m_0^2
\]
In the symmetric phase, $m_{0,\mathrm{ph}}^2 > 0$, and the action $S'$
describes $N$ free particles of mass $m_{0,\mathrm{ph}}$ in lattice
units.  Hence, the propagator is given by
\[
\vev{\phi^I_{\vec{n}} \phi^J_{\vec{0}}}_{S'}
 = \delta^{IJ} \int_{|k_i| < \pi} \frac{d^3 k}{(2\pi)^3} \,
 \frac{\e^{i k \cdot n}}{4 \sum_{i=1}^3 \sin^2 \frac{k_i}{2} +
 m_{0,\mathrm{ph}}^2 }
\]
This is a free field theory.  Denoting the physical length of a
lattice spacing as $a = \frac{1}{\mu \e^t}$, $m_{0,\mathrm{ph}}^2$ is
related to the physical squared mass $m_{\mathrm{ph}}^2$ as
\[
m_{0, \mathrm{ph}}^2 = \e^{- 2t} \frac{m_{\mathrm{ph}}^2}{\mu^2}
\stackrel{t \to \infty}{\longrightarrow} 0
\]

Thus, in the limit $N \to \infty$, the scalar fields are free.  The
interactions are due to the fluctuations of $\sigma_{\vec{n}}$, and
are of order $\frac{1}{N}$.

\subsubsection{Determination of $\Delta m_0^2$}

We now obtain $\Delta m_0^2$ as
\[
\Delta m_0^2 = \lambda_0 \vev{\frac{(\phi_{\vec{n}})^2}{2}}_{S'} =
\frac{\lambda_0}{2} \int_{|k_i|<\pi} \frac{d^3 k}{(2\pi)^3} \frac{1}{
4 \sum_i \sin^2 \frac{k_i}{2} + m_{0,\mathrm{ph}}^2}
\]
For $t \gg 1$, $m_{0,\mathrm{ph}}^2 = \frac{m_{\mathrm{ph}}^2}{\mu^2}
\e^{-2t}$ is small, and we can approximate the integral as
\begin{eqnarray*}
&&\int_{|k_i|<\pi} \frac{d^3 k}{(2\pi)^3} \frac{1}{ 4 \sum_i \sin^2
\frac{k_i}{2} + m_{0,\mathrm{ph}}^2} - \int_{|k_i|<\pi} \frac{d^3
k}{(2\pi)^3} \frac{1}{4 \sum_i \sin^2 \frac{k_i}{2}} \\ &=& -
m_{0,\mathrm{ph}}^2 \int_{|k_i|<\pi} \frac{d^3 k}{(2\pi)^3} \frac{1}{(
4 \sum_i \sin^2 \frac{k_i}{2} + m_{0,\mathrm{ph}}^2 ) \cdot 4 \sum_i
\sin^2 \frac{k_i}{2}} \\ &\stackrel{t \gg 1}{\Longrightarrow}& - \e^{-t}
\frac{m_{\mathrm{ph}}^2}{\mu} \int \frac{d^3 k}{(2\pi)^3} \frac{1}{k^2
(k^2 + m_{\mathrm{ph}}^2)} = - \frac{1}{4 \pi}
\frac{m_{\mathrm{ph}}}{\mu} \e^{-t}
\end{eqnarray*}
Therefore, we obtain
\[
\Delta m_0^2 = \frac{\lambda_0}{2} A_3  - \frac{\lambda_0}{8 \pi}
\frac{m_{\mathrm{ph}}}{\mu} \e^{-t}
\]
where
\[
A_3 \equiv \int_{|k_i|<\pi} \frac{d^3
k}{(2\pi)^3} \frac{1}{4 \sum_i \sin^2 \frac{k_i}{2}}
\]
is a constant.  Thus,
\[
m_{0,\mathrm{ph}}^2 = m_0^2 + \Delta
 m_0^2 = m_0^2 + \frac{\lambda_0}{2} A_3 - \frac{\lambda_0}{8 \pi}
 \frac{m_{\mathrm{ph}}}{\mu} \e^{-t}
\]
For large $t \gg 1$ we can ignore $m_{0,\mathrm{ph}}^2 \propto
\e^{-2t}$, and we obtain
\[
\ffbox{m_0^2 = - \frac{\lambda_0}{2} A_3 + \frac{\lambda_0}{8 \pi}
\frac{m_{\mathrm{ph}}}{\mu} \, \e^{-t}}
\]
Comparing this with the expected result
\[
m_0^2 = m_{0,cr}^2 (\lambda_0) + z_m (\lambda_0) \frac{g}{\mu^2} \e^{-
y_E t}
\]
we obtain
\[
\ffbox{m_{0,cr}^2 (\lambda_0) = - \frac{\lambda_0}{2} A_3}\quad \&\quad
\ffbox{y_E = 1} \quad\&\quad z_m (\lambda_0) = \frac{\lambda_0}{8 \pi}
\]
We also obtain
\[
\frac{g}{\mu^2} = \frac{m_{\mathrm{ph}}}{\mu}
 \longrightarrow m_{\mathrm{ph}} = \frac{g}{\mu} \propto g
\]
as expected from the general result $m_{\mathrm{ph}}\propto
g^{\frac{1}{y_E}}$.

Before closing this subsection, we remark that the above result cannot
possibly be obtained by perturbation theory.  The expression for the
physical squared mass can be written as
\[
m_{0, \mathrm{ph}}^2 = \left( \frac{8 \pi}{\lambda_0} m_0^2 +
\frac{2}{\pi} \right)^2
\]
which diverges as $\lambda_0 \to 0$.

\subsubsection{Continuum limit in the large $N$}

Thus, in the large $N$ limit, the continuum limit of the two-point
function is obtained as
\[
\vev{\phi^I (\vec{r}) \phi^J (\vec{0})}_g
\equiv \mu \lim_{t \to \infty} \e^t \vev{\phi^I_{\vec{n} = \vec{r}
 \mu \e^t} \phi^J_{\vec{0}}}_{m_0^2 = m_{0,cr}^2 (\lambda_0) +
 \frac{g}{\mu^2} \e^{-t}}
= \int \frac{d^3 p}{(2\pi)^3} \frac{\e^{i p r}}{p^2 +
 m_{\mathrm{ph}}^2}
\]
where
\[
 m_{\mathrm{ph}}^2 \equiv \left(\frac{8 \pi}{\lambda_0}\right)^2
 \, \frac{g^2}{\mu^2}
\]
This is a free theory, and the anomalous dimension vanishes:
$\ffbox{\eta \stackrel{N \to \infty}{\longrightarrow} 0}$

\subsubsection{Results from the $\ep$ expansions}

We have just seen that the theory becomes free in the limit $N \to
\infty$.  Similarly, if we define the theory in
\[
D \equiv 4 - \ep
\]
dimensional space ($\ep = 1$ for three dimensional space), the theory
is known to become trivial in the limit $\ep \to 0$.  The critical
exponents $y_E$, $\eta$ can then be calculated in powers of $\ep$, and
the following results are known\cite{Wilson:1973jj}:
\[
\left\lbrace
\begin{array}{c@{~=~}l}
 y_E & 2 - \frac{N+2}{N+8} \ep + \mathrm{O} (\ep^2, 1/N)\\ \eta &
 \frac{1}{2} \frac{N+2}{(N+8)^2} \ep^2 + \mathrm{O} (\ep^3, 1/N^2)
\end{array}
\right.
\]
For $N=1$, by substituting $\ep = 1$, we obtain
\[
y_E \simeq \frac{5}{3},\quad \eta \simeq 0.02
\]
The numerical simulations suggest $\eta \simeq 0.04$, and the leading
order approximation in $\ep$ is not a good fit, while $y_E \simeq
\frac{5}{3}$ is a good fit.

\subsection{$4$ dimensions}

Following the procedure given for the $3$ dimensional case above, we
can rewrite the action using an auxiliary field $\alpha_{\vec{n}}$ as
\[
S = \sum_{\vec{n}} \left( \frac{1}{2} \sum_{i=1}^4 \left(
\phi^I_{\vec{n}+\hat{i}} - \phi^I_{\vec{n}}\right)^2 + \frac{m_0^2}{2}
(\phi^I_{\vec{n}})^2 + \frac{N}{2 \lambda_0} \alpha_{\vec{n}}^2 - i
\alpha_{\vec{n}} \, \frac{(\phi^I_{\vec{n}})^2}{2} \right)
\]

From now on we assume the symmetric phase $m_0^2 > m_{0,cr}^2
(\lambda_0)$.  (We will compute the critical value shortly.)  We shift
the auxiliary field by the saddle point value
\[
\alpha_{\vec{n}} = i \Delta m_0^2 + \sqrt{\frac{\lambda_0}{N}}\,
\sigma_{\vec{n}}
\]
where we have rescaled the fluctuation $\sigma_{\vec{n}}$ for later
convenience.  The action is now written as
\begin{eqnarray*}
S &=& \sum_{\vec{n}} \Bigg[ \frac{1}{2} \sum_{i=1}^4 \left(
\phi^I_{\vec{n}+\hat{i}} - \phi^I_{\vec{n}} \right)^2 +
\frac{m_{0,\mathrm{ph}}^2}{2} (\phi^I_{\vec{n}})^2 \\
&& \quad + \frac{1}{2} \sigma_{\vec{n}}^2 - i
\sqrt{\frac{\lambda_0}{N}}\, \sigma_{\vec{n}} \frac{1}{2}
(\phi^I_{\vec{n}})^2 + i \sqrt{\frac{N}{\lambda_0}} \Delta m_0^2 \,
\sigma_{\vec{n}} \Bigg]
\end{eqnarray*}
where
\[
m_{0,\mathrm{ph}}^2 \equiv m_0^2 + \Delta m_0^2
\]

The value $i \Delta m_0^2$ of $\alpha_{\vec{n}}$ at the saddle point
is determined by the same condition as for the $3$ dimensional case
\[
\Delta m_0^2 = \frac{\lambda_0}{2} \int_{|k_i| < \pi} \frac{d^4
k}{(2\pi)^4} \frac{1}{4 \sum_{i=1}^4 \sin^2 \frac{k_i}{2} +
m_{0,\mathrm{ph}}^2}
\]
except that the momentum integral is now four dimensional.  We find,
for small $m_{0,\mathrm{ph}}^2$,
\[
\Delta m_0^2 = \frac{\lambda_0}{2} \left( A_4 +
m_{0,\mathrm{ph}}^2 \frac{1}{(4\pi)^2} (\ln m_{0,\mathrm{ph}}^2 +
B_4 ) \right)
\]
where $A_4$ is a constant defined by
\[
A_4 \equiv \int_{|k_i|<\pi} \frac{d^4 k}{(2\pi)^4} \frac{1}{4 \sum_i
\sin^2 \frac{k_i}{2}}
\]
and $B_4$ is a constant (unknown to me at least).  Therefore,
\[
m_{0,\mathrm{ph}}^2 = m_0^2 + \frac{\lambda_0}{2} \left( A_4 +
m_{0,\mathrm{ph}}^2 \frac{1}{(4\pi)^2} (\ln m_{0,\mathrm{ph}}^2 +
B_4 ) \right)
\]
For the continuum limit, we must obtain
\[
m_{0,\mathrm{ph}}^2 = \frac{m_{\mathrm{ph}}^2}{\mu^2} \e^{- 2t}
\stackrel{t \to \infty}{\longrightarrow} 0
\]
Hence, we obtain for large $t$
\begin{eqnarray*}
m_0^2 &=& - \frac{\lambda_0}{2} A_4 + \e^{- 2t}
\frac{m_{\mathrm{ph}}^2}{\mu^2} \left( 1 - \frac{\lambda_0}{2}
\frac{1}{(4\pi)^2} \left( - 2 t + \ln \frac{m_{\mathrm{ph}}^2}{\mu^2}
+ B_4\right) \right)\\ &=& - \frac{\lambda_0}{2} A_4 + \e^{-2t}
\frac{\lambda_0}{(4\pi)^2} \frac{m_{\mathrm{ph}}^2}{\mu^2} \left(
\frac{(4\pi)^2}{\lambda_0} + t - \frac{1}{2} \ln
\frac{m_{\mathrm{ph}}^2}{\mu^2} - \frac{B_4}{2} \right)
\end{eqnarray*}
Therefore, the critical squared mass is given by
\[
\ffbox{m_{0,cr}^2 (\lambda_0) = - \frac{\lambda_0}{2} A_4}
\]

We now introduce two parameters:
\[
\ffbox{
\begin{array}{c@{~\equiv~}l}
 \tilde{\lambda} & \frac{1}{t + \frac{(4\pi)^2}{\lambda_0} -
 \frac{B_4}{2}}\\
 \tilde{m}^2 & m_{\mathrm{ph}}^2 \left( 1 - \tilde{\lambda} \ln
 \frac{m_{\mathrm{ph}}}{\mu} \right)
\end{array}
}
\]
so that the above result can be written nicely as
\[
\ffbox{m_0^2 = m_{0,cr}^2 (\lambda_0) + z_m (\lambda_0) \e^{- 2t}
\tilde{\lambda}^{- 1} \frac{\tilde{m}^2}{\mu^2} }
\]
where
\[
z_m (\lambda_0) \equiv \frac{\lambda_0}{(4\pi)^2}
\]
We note 
\[
\left\lbrace
\begin{array}{c@{~\stackrel{t \to \infty}{\longrightarrow}~}c}
\tilde{\lambda} & 0\\
\tilde{m}^2 & m_{\mathrm{ph}}^2
\end{array}
\right.
\]

To find the physical meaning of $\tilde{\lambda}$, we must look at
interactions which are of order $\frac{1}{N}$.  The interactions are
mediated by the $\sigma$ field.  Let us look at the two-point function
of $\sigma$ to order $1$.  The self-energy correction is given by
\parbox{3cm}{\hfill$- \Pi (k)
\equiv\quad$}\parbox{4cm}{\epsfig{file=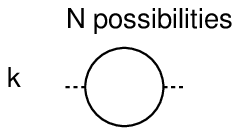}\hfill}
\[
= - \frac{\lambda_0}{2} \int_{|p_i|<\pi} \frac{d^4 p}{(2\pi)^4}
\frac{1}{\left( 4 \sum_i \sin^2 \frac{p_i}{2} + m_{0,\mathrm{ph}}^2
\right) \left( 4 \sum_j \sin^2 \frac{p_j + k_j}{2} + m_{0,\mathrm{ph}}^2
\right)}
\]
For $k^2 = \mathrm{O} (\e^{-2t})$ and $m_{0,\mathrm{ph}}^2 =
\mathrm{O} (\e^{-2t})$ this integral can be evaluated.  We find
\[
\Pi (k) = - \frac{\lambda_0}{2} \frac{1}{(4\pi)^2} \left(
\ln m_{0,\mathrm{ph}}^2 + B_4 
+ f \left(\frac{4 m_{0,\mathrm{ph}}^2}{k^2} \right) \right)
\]
where
\[
f(x) \equiv \sqrt{1 + x} \ln \frac{\sqrt{1+x}+1}{\sqrt{1+x}-1} - 1
\]

Thus, the propagator of $\sigma$ is given by
\[
\vev{\sigma_{\vec{n}} \sigma_{\vec{0}}} =
 \int_{|k_i|<\pi} \frac{d^3 k}{(2\pi)^3} \,\frac{\e^{i k n}}{1 + \Pi (k)}
\]
and in the momentum space the four-point vertex is obtained as\\
\parbox{5cm}{\hfill\epsfig{file=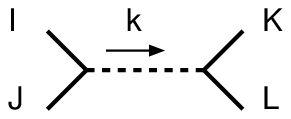}}\parbox{5cm}{
\[
= - \delta^{IJ} \delta^{KL} \frac{1}{N} \frac{\lambda_0}{1 + \Pi (k)}
\]}\\
Now, for
\[
k^2 = \frac{k_{\mathrm{ph}}^2}{\mu^2} \e^{-2t},\quad
m_{0,\mathrm{ph}}^2 = \frac{m_{\mathrm{ph}}^2}{\mu^2} \e^{-2t}
\]
we can rewrite
\begin{eqnarray*}
\frac{\lambda_0}{1 + \Pi (k)} &=& \frac{1}{\frac{1}{\lambda_0} -
\frac{1}{2} \frac{1}{(4\pi)^2} \left( \ln
\frac{m_{\mathrm{ph}}^2}{\mu^2} - 2 t + B_4 +
f\left(\frac{4 m_{\mathrm{ph}}^2}{k_{\mathrm{ph}}^2}\right) \right)}\\
&=& \frac{(4\pi)^2}{\frac{1}{\tilde{\lambda}} - \ln
\frac{m_{\mathrm{ph}}}{\mu} -
\frac{1}{2} f\left( \frac{4 m_{\mathrm{ph}}^2}{k_{\mathrm{ph}}^2}
\right) } \stackrel{\tilde{\lambda} \ll 1}{\longrightarrow} (4\pi)^2
\tilde{\lambda} 
\end{eqnarray*}
For small $\tilde{\lambda}$, the strength of the interaction is of
order $\tilde{\lambda}$.  Hence, we can call $\tilde{\lambda}$ a
coupling constant.  

\subsubsection{Real continuum limit}

Now, let us see what happens if we take the continuum limit $t \to
\infty$.  A disturbing thing happens.  Since
\[
\tilde{\lambda} = \frac{1}{t + \frac{(4\pi)^2}{\lambda_0} -
\frac{B_4}{2}} \stackrel{t \to \infty}{\longrightarrow} 0
\]
the interaction disappears in this limit, obtaining a free massive
theory of mass $\tilde{m}$.  Thus, we find
\[
\vev{\phi^I (\vec{r}) \phi^J (\vec{0})}_{\tilde{m}^2} \equiv \mu
 \lim_{t \to \infty} \e^{2 t} \vev{\phi^I_{\vec{n} = \vec{r} \mu \e^t}
 \phi^J_{\vec{0}}}_{m_0^2, \lambda_0} = \delta^{IJ} \int \frac{d^3
 k}{(2 \pi)^3} \frac{\e^{i k r}}{k^2 + \tilde{m}^2}
\]
where $m_0^2$ is given by
\[
m_0^2 = - \frac{\lambda_0}{2} A_4 + z_m (\lambda_0) \e^{-2t} \left( t
+ \frac{(4\pi)^2}{\lambda_0} - \frac{B_4}{2} \right)
\frac{\tilde{m}^2}{\mu^2} \stackrel{t \to \infty}{\longrightarrow} 0
\]
Note that for $t \gg 1$ the coefficient of $\tilde{m}^2$ behaves like
$t \e^{-2t}$, different from the simple $\e^{-2t}$, due to the
non-vanishing $\lambda_0$.  Nevertheless the non-vanishing $\lambda_0$
does not give rise to interactions in the continuum limit.

\subsubsection{Almost continuum limit}

The only way to keep the interaction non-vanishing is to keep $t$
\underline{large but finite}.  For large $t$ the coupling constant
$\tilde{\lambda}$ is small, and hence the smallness of
$\tilde{\lambda}$ is a sign that the space is pretty close to
continuum.  Given a coupling constant $\tilde{\lambda}$, the lattice
spacing is given by
\[
a = \frac{1}{\mu} \e^{-t} = \frac{1}{\mu} \exp \left( -
\frac{1}{\tilde{\lambda}} + \frac{(4\pi)^2}{\lambda_0} - \frac{B_4}{2}
\right) 
\]
in physical units.  By defining
\[
\ffbox{\Lambda_L (\tilde{\lambda}) \equiv \mu\,
\e^{\frac{1}{\tilde{\lambda}}}}
\]
we can write the above as
\[
\ffbox{a = \frac{1}{\Lambda_L (\tilde{\lambda})}\, \e^{\frac{(4
\pi)^2}{\lambda_0} - \frac{B_4}{2}}}
\]

We now define an almost continuum limit for $t \gg 1$ by
\[
\ffbox{\vev{\phi^I (\vec{r}) \phi^J (\vec{0})}_{\tilde{m}^2,
\tilde{\lambda};\, \mu} \equiv \mu^2 \e^{2 t}
\vev{\phi^I_{\vec{n}=\vec{r}\mu \e^t} \phi^J_{\vec{0}}}_{m_0^2,
\lambda_0}}
\]
where $\lambda_0$ is finite, and $m_0^2$ and $\tilde{\lambda}$ are
given as before by
\[
\ffbox{
\begin{array}{c@{~=~}l}
m_0^2 & m_{0,cr}^2 (\lambda_0) + z_m (\lambda_0) \e^{-2t}
\tilde{\lambda}^{-1} \frac{\tilde{m}^2}{\mu^2}\\
\tilde{\lambda} & \frac{1}{t + \frac{(4\pi)^2}{\lambda_0} -
\frac{B_4}{2}}
\end{array}
}
\]
The almost continuum limit depends on $t$ through $\tilde{\lambda}$.

\subsubsection{RG equations}

Using the definition of the almost continuum limit, we can derive RG
equations.  Given $m_0^2, \lambda_0$, we wish to find the changes in
$\tilde{m}^2, \tilde{\lambda}$ as we change $t$ to $t + \Delta t$,
where $\Delta t$ is infinitesimal.  First look at $\tilde{\lambda}$:
\[
\tilde{\lambda} \stackrel{t \to t + \Delta t}{\longrightarrow}
 \frac{1}{t + \Delta t + \frac{(4\pi)^2}{\lambda_0} - \frac{B_4}{2}}
 = \tilde{\lambda} - \Delta t \cdot \tilde{\lambda}^2
\]
Then, to keep $m_0^2$ invariant, we must find
\[
\tilde{m}^2 \stackrel{t \to t + \Delta t}{\longrightarrow}
\e^{2 \Delta t} \frac{\tilde{\lambda} - \Delta t \cdot
\tilde{\lambda}^2}{\tilde{\lambda}} \tilde{m}^2
= \e^{2 \Delta t} ( 1 - \Delta t \cdot \tilde{\lambda} ) \tilde{m}^2
\]
The above changes of $\tilde{\lambda}, \tilde{m}^2$ give the RG
transformation of the parameters.  Then, the RG equation of the
two-point function are obtained as
\[
\ffbox{ \vev{\phi^I (\vec{r} \e^{- \Delta t}) \phi^J (\vec{0})}_{\e^{2 \Delta
t} \tilde{m}^2 ( 1 - \Delta t \cdot \tilde{\lambda} ), \tilde{\lambda}
- \Delta t \cdot \tilde{\lambda}^2; \, \mu}
 = \e^{2 \Delta t} \vev{\phi^I (\vec{r}) \phi^J
(\vec{0})}_{\tilde{m}^2, \tilde{\lambda};\, \mu} }
\]

To solve this RG equation, we first note that
\[
\Lambda_L (\tilde{\lambda} - \Delta t \cdot \tilde{\lambda}^2)
 = \e^{\Delta t} \Lambda_L (\tilde{\lambda})
\]
and then note that
\[
\frac{\tilde{m}^2}{\tilde{\lambda}}
\stackrel{\mathrm{RG}}{\longrightarrow}
\frac{\e^{2 \Delta t} \tilde{m}^2 (1 - \Delta t \cdot
\tilde{\lambda})}{\tilde{\lambda} - \Delta t \cdot \tilde{\lambda}^2}
 = \e^{2 \Delta t} \frac{\tilde{m}^2}{\tilde{\lambda}}
\]
Hence, the general solution of the RG equation is given by
\[
\vev{\phi^I (\vec{r}) \phi^J (\vec{0})}_{\tilde{m}^2,
\tilde{\lambda};\, \mu}
= \delta^{IJ} C \left( \Lambda_L (\tilde{\lambda}) r,\,
\frac{\tilde{m}^2}{\tilde{\lambda} \Lambda_L (\tilde{\lambda})^2}\right)
\]
where $C$ is an arbitrary function of a single variable.

This implies that the physical mass is given in the form
\[
m_{\mathrm{ph}}^2 = \frac{\tilde{m}^2}{\tilde{\lambda}} f \left(
\frac{\tilde{m}^2}{\tilde{\lambda} \Lambda_L (\tilde{\lambda})^2}\right)
\]
or equivalently
\[
\tilde{m}^2 = \tilde{\lambda} m_{\mathrm{ph}}^2 \cdot g \left(
\frac{m_{\mathrm{ph}}^2}{\Lambda_L (\tilde{\lambda})^2}\right)
\]
This is consistent with the result we've already obtained:
\[
\tilde{m}^2 = m_{\mathrm{ph}}^2 \left( 1 - \tilde{\lambda} \ln
m_{\mathrm{ph}} \right) = \tilde{\lambda} m_{\mathrm{ph}}^2 \left(
\frac{1}{\tilde{\lambda}} - \ln m_{\mathrm{ph}} \right)
 = \tilde{\lambda} m_{\mathrm{ph}}^2 \cdot \frac{- 1}{2} \ln 
\frac{m_{\mathrm{ph}}^2}{\Lambda_L (\tilde{\lambda})^2}
\]
Thus, $g(x) = - \frac{1}{2} \ln x$.

\end{document}